\documentclass{aa}
\usepackage{graphics}

\begin{document}

\thesaurus{08 (02.13.5; 09.01.1; 09.03.1; 09.13.2; 09.09.1 CrA, Cha I)}
	
\titlerunning{Molecules in protostellar cloud cores}
\authorrunning{Kontinen et al.} 

\title{Molecular line study of evolution 
in protostellar cloud cores}

\author{S.~Kontinen\inst{1}, J.~Harju\inst{1}, A. Heikkil\"a\inst{1} and 
L.K. Haikala\inst{1,2}}

\offprints{S.~Kontinen}

\institute{Observatory, P.O. Box 14, FIN-000\,14 University of Helsinki, Finland
\and       Swedish-ESO Submillimetre Telescope, European Southern Observatory, 
           Casilla 19001, Santiago, Chile}
	   
\date{Received 11.04.2000; accepted 14.07.2000}

\maketitle

\begin{abstract}

Two dense dark cloud cores representing different stages of dynamical
evolution were observed in a number of molecular spectral lines. One
of the cores, Cha-MMS1 in the Chamaeleon \ion{cloud}{i} contains a
Class 0 protostar, whereas the other, CrA C in the R Coronae Australis
cloud, is pre-stellar.  The molecules selected for this study are
supposed to show significant abundance variations in the course of the
chemical evolution.

We find that the cores have very different chemical compositions.
Cha-MMS1 exhibits characteristics of so-called `early-type' chemistry
with high abundances of carbon-chain molecules such as HC$_3$N,
CH$_3$CCH and c-C$_3$H$_2$. However, it also has a large N$_2$H$^+$
abundance, which is expected only to build up at later stages. In
contrast, none of the carbon-chain molecules were detected in CrA
C. On the other hand, CrA C has a higher SO abundance than Cha-MMS1,
which according to chemistry models implies that it is chemically
`older' than Cha-MMS1. The most striking difference between the two
cores is seen in the HC$_3$N/SO abundance ratio, which is at least
three orders of magnitude higher in Cha-MMS1 than in CrA C. This
result is somewhat surprising since starless cores are usually
thought to be chemically younger than star-forming cores.

Because of the high N$_2$H$^+$ abundance, we suggest that Cha-MMS1
represents the `late-time cyanopolyyne peak' that is predicted to
occur when heavy molecules start to freeze onto grain surfaces
(\cite{ruffle97}). This would also be a more natural explanation for
the carbon-chain molecules than the `early-time' picture in view of
the fact that the core is presently collapsing to form a star. The
abundances observed in CrA C can be explained either by pure gas-phase
models at late stages of evolution, or by the `SO peak' which follows
the second cyanopolyyne peak (\cite{ruffle99}).  Thus, the dynamical
evolution in CrA C seems to have been very slow compared with that of
Cha-MMS1, and we discuss possible reasons for this.

We detected two SO emission maxima around Cha-MMS1, which lie
symmetrically on both sides of the core, approximately on the line
connecting the centre of Cha-MMS1 and the position of Herbig-Haro
object HH49/50.  These SO peaks may signify the lobes of a bipolar
outflow, and the observation supports the suggestion by
\cite{reipurth} that Cha-MMS1 is the central source of HH49/50.


\end{abstract}

\section{Introduction}

To understand the first steps of the star-formation process, we must
be able to follow the evolution of molecular cloud cores.  One
approach is to conduct spectral-line observations of molecular species
whose abundances vary significantly with time.  Conversely, combining
such observations with other types of diagnostics of cloud evolution,
the chemical models may be tested.

Examples of species that are believed to be useful for the
aforementioned purposes are complex hydrocarbons and
cyanopolyynes\footnote{Molecules with the generic form HC$_{(2n+1)}$N
where $n=1,2, \cdots$}. Their abundances are expected to peak at early
stages of the evolution of dark clouds (e.g. \cite{herbst89}).  This
is because their production requires a high abundance of free neutral
carbon (\ion{C}{i}); a condition that is satisfied at early times in
the chemical evolution, whereas at later stages practically all free
carbon has been exhausted in the production of CO. In contrast,
molecules such as NH$_3$ and SO increase in abundance with time,
because their formation involves relatively slow neutral-neutral
reactions (\cite{suzuki}; \cite{hirahara95};
\cite{nilsson}), or because they (e.g. SO) are destroyed by
\ion{C}{i} or C$^+$. 

The validity of these model predictions have, to date, been tested in
a rather limited sample of dark cloud cores. Most observations have
been confined to the Taurus Molecular Cloud 1 (TMC-1), L134A, and to a
number of ``Myers' cores'' (\cite{pratap}; \cite{swade};
\cite{myers83}, \cite{benson98}; for a review see \cite{dishoeck}).  
However, for current chemistry models to be rigorously tested, the
cloud sample needs to be more representative of the
ISM.  In the present paper we extend the sample of well-studied
objects by two potentially very young molecular cloud cores.

The first core selected for this study is located in the region of the
reflection nebula Cederblad 110 in the Chamaeleon dark \ion{cloud}{i}.
\cite{reipurth} detected a strong 1.3mm dust continuum
source towards this core, and designated it as Cha-MMS1. 
The core belongs to an extensive,
northeast--southwest directed filament of dense gas near the centre of
this cloud (\cite{mattila}).  The second core, CrA~C (\cite{harju93})
is located in the southeastern `tail' of the R~Coronae Australis
molecular cloud.  The distances to the \ion{Chamaeleon}{i} and the
R~Coronae Australis clouds are 150 and 170 pc, respectively
(\cite{knude}). {The smallest diameters of the C$^{18}$O cores are 
0.11 and 0.05 pc for CrA~C and Cha-MMS1, respectively. The C$^{18}$O 
column densities towards both cores are similar: $N({\rm C^{18}O}) 
\sim 2 \,10^{15} ~{\rm cm}^{-2}$, which corresponds to a molecular 
hydrogen column density of $N({\rm H_2}) \sim 10^{22} ~{\rm cm}^{-2}$ 
(\cite{harju93}; \cite{mattila}). The average H$_2$ densities derived
from these numbers assuming prolate geometries are $\sim 3 \, 10^4$ and 
$\sim 7 \, 10^4$ cm$^{-3}$ for CrA~C and Cha-MMS1, respectively. The 
kinetic temperature in CrA~C is 10 K or less (\cite{harju93}), whereas 
in Cha-MMS1 the gas kinetic temperature as derived from ammonia 
observations is slightly elevated with respect to the typical dark 
cloud value, being $\sim 14$~K (Toth et al. 2000, in preparation).}

Even though both cores are supposed to be young, they represent
different stages of cloud evolution. Cha-MMS1 probably contains a
Class 0 protostar (\cite{reipurth}), which has been recently
identified with a far-infrared source, designated Ced110 IRS~10, by
\cite{lehtinen}. The core lies next to a small group of newly-born stars, the
closest being Ced 110 IRS~4 (\cite{prusti}) and the newly detected
IRS~11 (\cite{lehtinen}), which belong to the infrared source Class
I. The core is associated with a compact, north-south oriented
molecular outflow (\cite{mattila}; \cite{prusti}; \cite{lehtinen}).
\cite{reipurth} suggested that Cha-MMS1 contains the driving source of
the outflow and the Herbig-Haro objects HH 49/50 located some $10\arcmin$
south of the core.  

CrA~C shows no sign of star formation activity.  The core
appears to be surrounded by an extensive halo: it is located near the
centre of a neutral atomic hydrogen ring with a diameter of
$40^\prime$ (\cite{llewellyn}) which may represent the outer boundary
of a massive \ion{H}{i} to H$_2$ conversion region.  Observations of
continuum emission at 100\,$\mu$m and 200\,$\mu$m wavelenghts with the
ISOPHOT instrument on the Infrared Space Observatory (ISO) reveal a
200$\mu$m excess coinciding with the centre of CrA~C (Heikkil\"a et
al. 2000, in preparation), suggesting that a very cold clump is
located inside this core.

Based on these facts, one could expect CrA~C to exhibit the
characteristics of `early time' chemistry, e.g. high abundances of
unsaturated carbon compounds, whereas Cha-MMS1 could be thought to
have reached a more evolved stage. In order to test these ideas, we
have estimated for these dense cores the column densities of several
molecules which show significant time
dependence in chemistry models. 

The selected molecules are discussed in the next Section.  In
Sect.~3 we describe the observational procedure. In Sect.~4 the
results from the observations are presented and these are further
discussed in Sect.~5. Finally, in Sect.~6 we summarize our
conclusions.

\section{Observed molecules}

We discuss here the spectroscopic and chemical properties of the
molecules included in this study. Table~\ref{table:molecules} lists
their chemical formulae, structure types (hfs means that the lines
have hyperfine structure; o/p and $A/E$ that the molecule has ortho
and para or $A$-symmetry and $E$-symmetry forms, respectively),
permanent electric dipole moments ($\mu_{\rm electric}$) along the
relevant molecular axises, and rotational constants ($A$, $B$ and
$C$). The data were extracted mainly from the JPL
catalogue\footnote{An updated version is available on the WWW at
http://spec.jpl.nasa.gov.} (\cite{poynter}); exceptions are the
CH$_3$OH data (\cite{sastry}; \cite{anderson90}) and the value for the
dipole moment of CCS (\cite{murukami}).  \\

\subsection{Spectroscopic properties}

\noindent
Carbon monosulphide, {\bf CS}, is a linear molecule with a simple
rotational spectrum; the states are denoted by the total angular
momentum quantum number $J$.  The simplicity of the spectrum is due to
the facts that in the electronic ground state, the electrons in the CS
molecule have paired spins and possess no orbital angular momentum.
Hence, the ground state is denoted $^1\Sigma$. In addition, the C and S
nuclei have no spin. \\

\noindent
The dicarbon sulphide radical, {\bf CCS}, and the sulphur monoxide
radical, {\bf SO}, have more complicated spectra than CS.  The 
electrons in the electronic ground states of CCS and SO have no
net orbital angular momentum, but the total spin is non-zero
due to two electrons having unpaired spins.  Thus, the spin quantum
number is $S=1$ and the electronic ground state is denoted $^3\Sigma$
(e.g. \cite{gordy}, Sect. 4.2).  The electron spin is coupled to the
weak magnetic field arising from the rotation of the molecule.  Since
$S$=1, for each molecular rotational level with the rotational angular
momentum quantum number $N>0$, the total angular momentum quantum
number, $J$, has three possible values: $J=N-1$, $J=N$ or $J=N+1$.
Consequently, the rotational levels $N>0$ are split into triplets,
whereas for $N$=0 only one state ($J_N=1_0$) exists
(e.g. \cite{yamamoto}). \\
%(Note that the molecular rotational angular momentum is usually
%designated by $J$ also when $S=0$, although the use of $N$ would be
%more logical in this case).   \\

\noindent
Hydrogen isocyanide, {\bf HNC}, is a linear molecule.  The nitrogen
nucleus has a non-zero spin, and therefore possesses an electric
quadrupole moment. This interacts with the electric field
gradient of the molecule, which depends on the rotation. Consequently,
the nuclear spin ($I=1$) is coupled to the molecular rotational
angular momentum, $J$, to yield the total angular momentum, $F$
(\cite{gordy}, Sect.~9.4). The resulting hyperfine splitting of the
rotational lines of HNC is smaller than for its isomer HCN. For
example, for the lowest rotational transition ($J=1\rightarrow0$) the
$F=1\rightarrow1$ and $F=0\rightarrow1$ components are separated from
the main component $F=2\rightarrow1$ by $-0.27$ and $+0.41$
km\,s$^{-1}$ respectively (\cite{frerking}).  Hence, 
the hyperfine structure is usually not resolved in observations due to Doppler
broadening in the source and the insufficient spectral resolution of the
spectrometers available.  \\

\noindent
Cyanoacetylene, {\bf HC$_3$N}, is also a linear molecule.  Due to its
high moment of inertia, it has a small rotational constant ($B$) and
its rotational transitions $J\rightarrow J-1$ lie relatively closely
in frequency. Many of these lines are located in the commonly observed
frequency bands, and it is often used in multi-transition studies.
Like HNC, it has hyperfine structure due to the nuclear spin of
nitrogen. For higher $J$-values, however, the relative energy differences
between different $F$-levels are very small, and so no
hyperfine structure is seen in the spectra observed here.  \\

\noindent
Methyl acetylene, {\bf CH$_3$CCH}, is a low dipole moment, symmetric
top molecule. Its energy levels are described by two quantum numbers
$J$ and $K$. The former represents the total angular momentum and the
latter is the projection of $J$ on the symmetry axis of the molecule.
The lowest energy state in each $K$-ladder is $J$=$K$.  CH$_3$CCH
exists in two chemically different forms, the so called $A$ and $E$
symmetry species, depending on the relative orientations of the spins
of the three hydrogen nuclei in the CH$_3$ group. For the $A$-symmetry
species, the $K$ quantum numbers are multiples of 3 (i.e. $K=3\,n$;
$n$=0,1,2,...), while the $E$-symmetry species has $K=3\,n+1$ and
$3\,n+2$; $n$=0,1,2,...\, .  The ground-state energies for $A$ and $E$
CH$_3$CCH differ by 8.02\,K, $E$ lying higher in energy.  There are no
allowed electric dipole transitions between two states belonging to
different $K$-ladders. Consequently, the relative populations of two
$K$-ladders are determined by collisional transitions and therefore
depend on the kinetic temperature.  In fact, the so-called `rotational
temperature', which can be determined from a single observation of a
$J\rightarrow J-1, K-$multiplet (the lines lie closely in frequency),
is considered to be a good approximation of the kinetic temperature
(e.g. \cite{bergin94}).  \\

\noindent
Cyclo-propenylidene, {\bf c-C$_3$H$_2$}, is one of the few cyclic
molecules detected in interstellar space. It is a slightly asymmetric,
almost oblate top.  C$_3$H$_2$ has {\sl ortho}- and {\sl para}-states
originating from the two possible relative orientations of the spins
of the two hydrogen nuclei (having $I=\frac{1}{2}$ each). In the 
ortho-state the spins are parallel ($I=1$), whereas in the para-state the
spins are anti-parallel ($I=0$).  Since the ortho- and para-forms
cannot be interchanged in electric dipole transitions or easily in
chemical reactions, they can be seen as two different chemical
compounds.  The ground-state of the ortho-form ($J_{K_a,K_c}=1_{0,1}$)
lies 2.35\,K higher than that of the para-form
($J_{K_a,K_c}=0_{0,0}$).  The electric dipole transitions of
c-C$_3$H$_2$ are of $b$-type.  This means that the quantum numbers
describing the projection of $J$ on the molecular $a-$axes and
$c-$axes, $K_a$ and $K_c$, change according to the rule $\Delta K_a =
\pm 1$, $\Delta K_c = \pm 1$ (see
\cite{vrtilek}). \\

\noindent
Diazenylium, {\bf N$_2$H$^+$}, is a linear molecular ion.  Its
rotational transitions have hyperfine structure due to the coupling of
the spins of the two nitrogen nuclei ($I_1$,$I_2$) to the rotational
angular momentum ($J$). The energy levels are labelled using the
quantum numbers $J$, $F_1$ and $F$, where $F_1 = J+I_1$ and $F = F_1 +
I_2$ (e.g. \cite{caselli95}; \cite{gordy}, Sect. 9.5).  The
$J=1\rightarrow 0$ transition observed here has seven well resolved
hyperfine components. \\

\noindent 
Methanol, {\bf CH$_3$OH}, is a slightly asymmetric top
(e.g. \cite{townes}, chapters 4 \& 12).  It has a rather complicated
rotational spectrum, since it exhibits so called hindered internal
rotation: the hydroxyl (OH) group can rotate with respect to the
methyl (CH$_3$) group.  This gives rise to two different forms of
CH$_3$OH between which both radiative and collisional transitions are
forbidden: the $A$-symmetry and $E$-symmetry forms.  Chemical
reactions that might convert $A$ to $E$ or vice versa are believed to
occur only on very long time scales.  Therefore the two forms of
CH$_3$OH can be treated as chemically distinct species. The
ground-state of the $E$-symmetry form ($J_k=1_{-1}\, E$) lies 6.95\,K
higher than that of the $A$-symmetry form ($J_k=0_0\, A^+$). The
relevant quantum numbers describing the rotational spectrum are $J$
and $k$, with $k=K_a-K_c$. The electric dipole transitions observed
here are of $a$-type, i.e. they obey the selection rule $\Delta k=0$.
Like other symmetric or slightly asymmetric tops, CH$_3$OH may provide
useful information about the gas temperature.

\subsection{Chemistry}

The buildup of cyanopolyynes proceeds mainly via reactions between
complex hydrocarbon ions and nitrogen atoms (\cite{herbst89}).
HC$_3$N is the simplest cyanopolyyne, and its abundance correlates
well with heavier cyanopolyynes (e.g. \cite{federman}).  Due to its
large dipole moment, as well as being chemically an early-time
species, HC$_3$N should be a useful probe of the conditions in
pre-stellar cloud cores. \\

The main reactions leading to HC$_3$N and CH$_3$CCH involve either the
C$_3$H$_2^+$ or C$_3$H$_3^+$ ions (\cite{huntress}). The neutral
reactions involving C$_2$H$_2$ and CCH suggested recently by
\cite{turner98} and \cite{turner99} do not change this fact, since
these species are also derivatives of the aforementioned molecular ions.
Therefore, HC$_3$N and CH$_3$CCH abundances are expected to be well
correlated. \\

Reactions of nitrogen-bearing ions, such as HCNH$^+$, with neutral
hydrocarbons may also contribute to the production of cyanopolyynes
(\cite{herbst89}).  Since HCNH$^+$ is the precursor ion of HCN and HNC
(\cite{herbst78}), a correlation between these isomeric molecules and
cyanopolyynes is possible. In this paper, we shall not address the as
yet unresolved problem of the variation of the HNC/HCN abundance ratio
from source to source (for a recent overview see
e.g. \cite{talbi98}). We instead used HNC merely to survey the
extension of the core CrA~C, where HC$_3$N was not detected. This
should be as abundant as HCN but easier to detect due to the smaller
hyperfine splitting, \\

In their survey towards dense cores of dark clouds, \cite{suzuki} found
a strong correlation between the CCS and HC$_3$N column densities,
and, using data from \cite{cox}, a weaker correlation between CCS and
C$_3$H$_2$. They explained these correlations by the production of 
CCS in reactions between hydrocarbons and S$^+$, which should be abundant in
regions where also carbon-chain molecules are formed (\cite{prasad}).
The formation of CS also involves S$^+$ and occurs in the very early
stages of chemical evolution.  Thereafter, CS is mainly recycled via
the thioformyl ion, HCS$^+$, and its abundance is considered to be
rather constant in time (\cite{nejad}). \\

The correlation between HC$_3$N and CCS is due to fact that C$_3$H
dominates the production of both molecules.  HC$_3$N is related to
C$_3$H$_2$ via the neutral-neutral reaction ${\rm C_3H_2} + {\rm N}
\rightarrow {\rm HC_3N} + {\rm H}$ (\cite{herbst89}; \cite{nejad}). 
On the other hand, complex carbon chain molecules and CCS show no
correlation with ammonia, NH$_3$ (e.g. \cite{little}; \cite{suzuki}).
This can be explained by the fact that the production of NH$_3$
involves molecular nitrogen, N$_2$, which is formed in neutral-neutral
reactions and becomes abundant in the later stages of cloud evolution
when carbon chain molecules have already been lost in reactions with
ions such as He$^+$, H$^+$ and H$_3^+$ (\cite{suzuki}).
N$_2$H$^+$ is also formed from N$_2$ by ${\rm N_2} + {\rm H_3^+}
\rightarrow {\rm N_2\rm H^+} + {\rm H_2}$ and its abundance has been
found to follow closely that of NH$_3$ (\cite{hirahara95};
\cite{nejad}). Therefore, if the CCS/NH$_3$ abundance ratio can be
used as an age indicator as suggested by \cite{suzuki}, the same
should be true for CCS/N$_2$H$^+$. The depletion of CO onto grain
surfaces further increases the N$_2$H$^+$ abundance by making H$_3^+$
available for non-carbon bearing species to react with
(\cite{nejad}). \\

SO is destroyed by neutral carbon and, like N$_2$H$^+$, it is formed
in neutral-neutral reactions and benefits from the freezing-out of CO.
Its abundance therefore increases slowly and probably
remains low as long as the gas is rich in neutral atomic carbon. The
SO abundance first increases when carbon is consumed in the production
of CO, i.e. at late stages in the chemical evolution (\cite{bergin97};
\cite{nejad}). The late peaking of SO and the stability of CS (until
it too is frozen out) has led to the suggestion that the SO/CS
abundance could be used as a chemical clock
(\cite{ruffle99}). \cite{nilsson} came to the same conclusion on the
basis of pure gas phase chemistry models. However, the SO/CS ratio is
also sensitive to the gas-phase O/C ratio. Thus, when observing the
SO/CS ratio in two individual clouds, it is not easy to disentangle the
effects of time evolution from those of the O/C ratio.  \\

The precursor ion of CH$_3$OH, CH$_3$OH$_2^+$, is formed from the
radiative association reaction between CH$_3^+$ and H$_2$O
(e.g. \cite{herbst89}).  The close relation to CH$_3^+$ makes CH$_3$OH
an early-time molecule, as long as gas phase reactions are
considered. The large abundances of CH$_3$OH in high-mass star-forming
cores has been explained by evaporation from grain surfaces due to
shock heating (e.g. \cite{menten}). \\

\section{Observations}

The observations were made in February 1998 at the Swedish-ESO
Submillimetre Telescope (SEST) on La Silla in Chile. The 100 and 150
GHz SIS receivers were used simultaneously. These were connected to a
2000 channel acousto-optical spectrometer (AOS), which was split in
two bands of 43 MHz each.  The resulting channel-separations are
0.13\,km\,s$^{-1}$ (at 100\,GHz) and 0.09\,km\,s$^{-1}$ (at 150\,GHz),
respectively.  Calibration was achieved by the chopper-wheel
method. The pointing and focus of the telescope were checked at 3-4
hours intervals against the SiO($v=1,J=2\rightarrow1$) maser line towards
AH~Sco, U~Men and R~Dor.  The pointing accuracy was typically
$3\arcsec$ (rms) in each coordinate. 

Most of the lines were observed in the frequency switching mode with a
frequency throw of 6 MHz. Position switching was used for CH$_3$CCH and
N$_2$H$^+$ since the spectrometer band covers several of their lines. 

The observations were started by mapping both sources in the
HC$_3$N($J=10\rightarrow 9$) and SO($J_N=4_3\rightarrow 3_2$)
transitions simultaneously in the frequency switching mode. 
HC$_3$N was not detected in CrA~C, and therefore the map was repeated in the
transition pair HNC($J=1\rightarrow 0$) \& SO($J_N=4_3\rightarrow
3_2$).  The rest of the observations were single position pointings
towards the SO (in CrA~C) or HC$_3$N (in Cha-MMS1) line emission peaks
in these maps.  Table~\ref{table:coordinates} lists the $(0,0)$ 
positions used in the mapping observations
and the peak positions, towards which the single pointing observations
were made in each core. \\

In Table~\ref{table:lines} we give a summary of the observed lines and
some telescope characteristics at the corresponding frequencies.  The
columns of this Table are: (1) the rest-frequency; (2) the molecule;
(3) the transition; (4) the line strength of the transition
($S_{ul}$); (5) the energy of the upper level of the transition
($E_{u}/k$); (6) and (7) the half-power beam-width ($\Theta_{\rm mb}$)
and the main beam efficiency ($\eta_{\rm mb}$) of the antenna.
Further details of the SEST are available e.g. on the WWW at
http://www.ls.eso.org/lasilla/Telescopes/SEST.

\section{Results}

\subsection{Maps}

\begin{figure*} 
\resizebox{12cm}{!}{\includegraphics{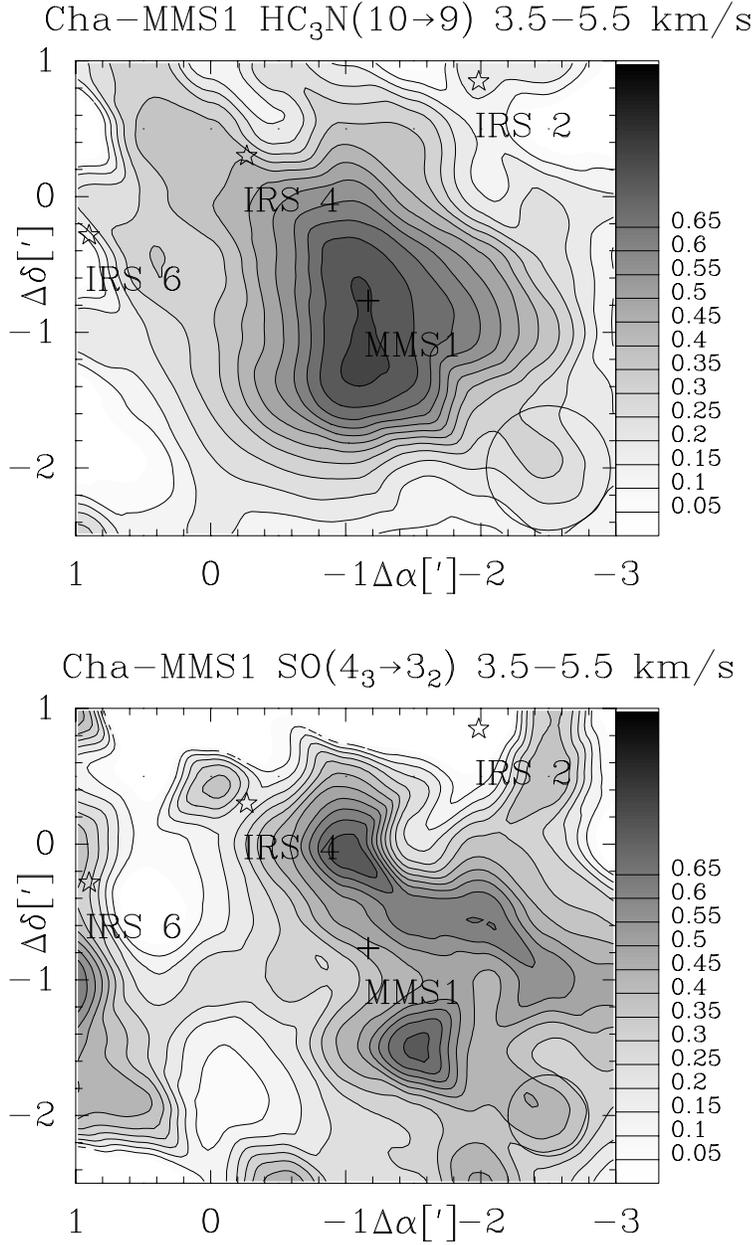}}
\caption[]{The HC$_3$N($J=10\rightarrow9$) and SO($J_N = 4_3
\rightarrow 3_2$) integrated intensity ($\int T_{\rm A}^*\,dv$) maps of Cha-MMS1. 
The velocity range and the intensity levels (in K\,km\,s$^{-1}$) are
indicated in the
Figure. Typical $3\sigma$ rms noise levels are 0.04 K\,km\,s$^{-1}$
and 0.11\,K\,km\,s$^{-1}$ in the HC$_3$N and the SO maps,
respectively.  The locations of the 1.3 mm dust continuum peak MMS1
(\cite{reipurth}), and the nearest embedded infrared sources
(\cite{prusti}) are denoted by a cross and stars, respectively. The
beamsizes are indicated in the bottom right of each map. The
coordinates of the $(0,0)$ position are given in
Table~\ref{table:coordinates}.}
\label{figure:chamaps} 
\end{figure*}

The velocity-integrated HC$_3$N($J=10\rightarrow 9$) and SO($J_N =
4_3\rightarrow 3_2$) line-intensity maps of Cha-MMS1 are presented in
Fig.~\ref{figure:chamaps}. The locations of the embedded infrared 
sources Ced 110 IRS2, IRS4 and IRS6 (\cite{prusti}), and the peak 
position of the 1.3 mm radio continuum source Cha-MMS1
(\cite{reipurth}) are also indicated. It is evident from the map that the
HC$_3$N distribution is compact and peaks close to the dust continuum
maximum, some $30\arcsec$ south of it. A comparison with Fig.~1 of
\cite{reipurth} shows that the dust continuum emission and the HC$_3$N
line emission have similar sizes and shapes, and therefore conceivably
originate from the same source, despite the slight shift in the peak
positions.  The infrared point sources lie near the northeastern edge
of the molecular cloud core.  SO line emission is present in the
compact core, but shows no clear concentration.  The maxima lie on its
northern and southern sides, and are possibly related to a molecular
outflow (see Sect. 5). \\

\begin{figure*} 
\resizebox{12cm}{!}{\includegraphics{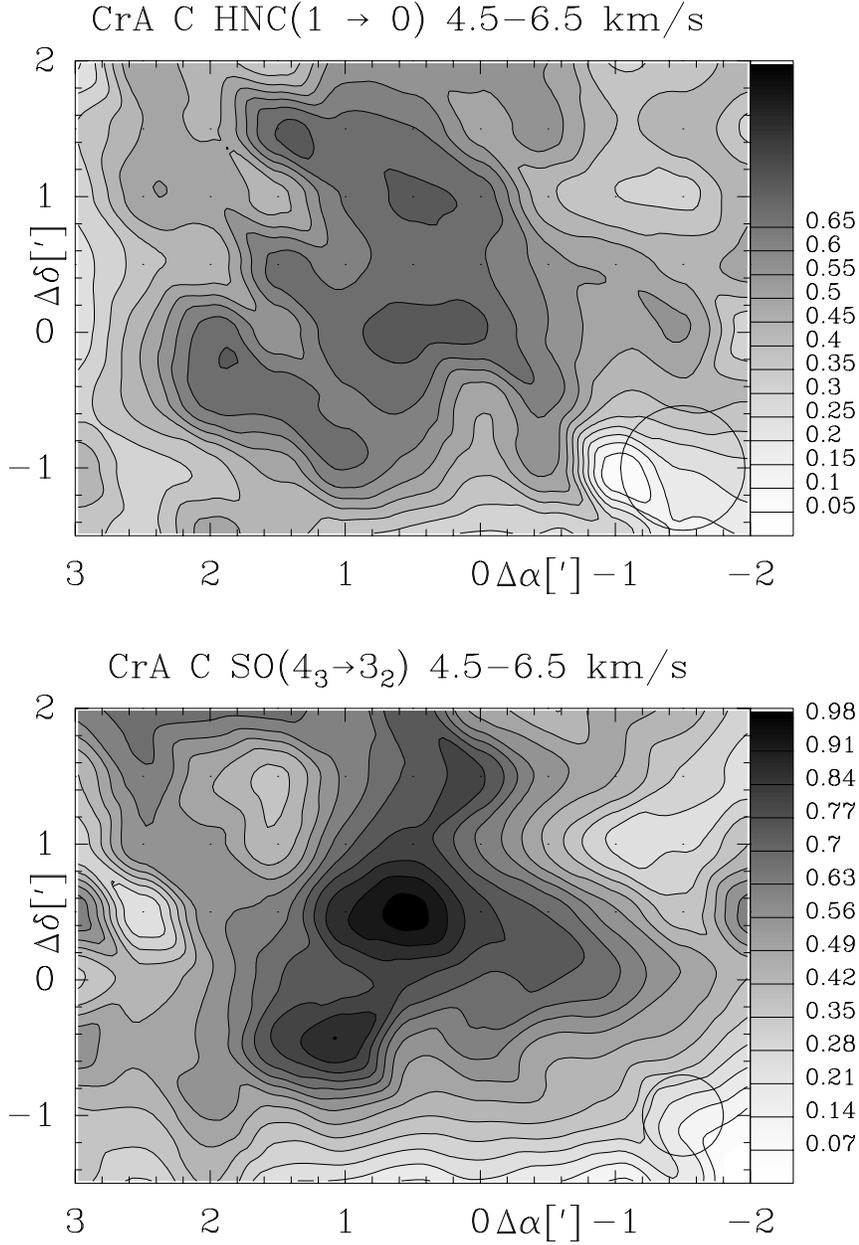}}
\caption[]{The HNC($J=1\rightarrow0$) and SO($J_N = 4_3
\rightarrow 3_2$) integrated intensity ($\int T_{\rm A}^*\,dv$) maps
of CrA~C. The velocity range and the intensity levels (in K\,km\,s$^{-1}$) are
indicated in the
Figure. Typical $3\sigma$ rms noise levels are 
0.06 K\,km\,s$^{-1}$ and 0.11\,K\,km\,s$^{-1}$ in the HNC and the SO
maps, respectively. The beamsizes are indicated in the bottom right of
each map, and the coordinates of the $(0,0)$ position are given in
Table~\ref{table:coordinates}.}
\label{figure:cramaps} 
\end{figure*}

The integrated HNC($J=1\rightarrow0$) and SO($J_N = 4_3 \rightarrow
3_2$) intensity maps of CrA~C are shown in
Fig.~\ref{figure:cramaps}. In these maps, the situation is the
opposite of that in the maps of the Cha-MMS1 core. HNC, which was
chosen instead of the undetected HC$_3$N, is weak and dispersed,
whereas SO has a strong peak towards the offset
$(30\arcsec,30\arcsec)$ from the centre of the map, and a weaker
second maximum towards the offset $(60\arcsec,-30\arcsec)$.

\subsection{Spectra and line parameters}

\begin{figure*} 
\resizebox{12cm}{!}{\includegraphics{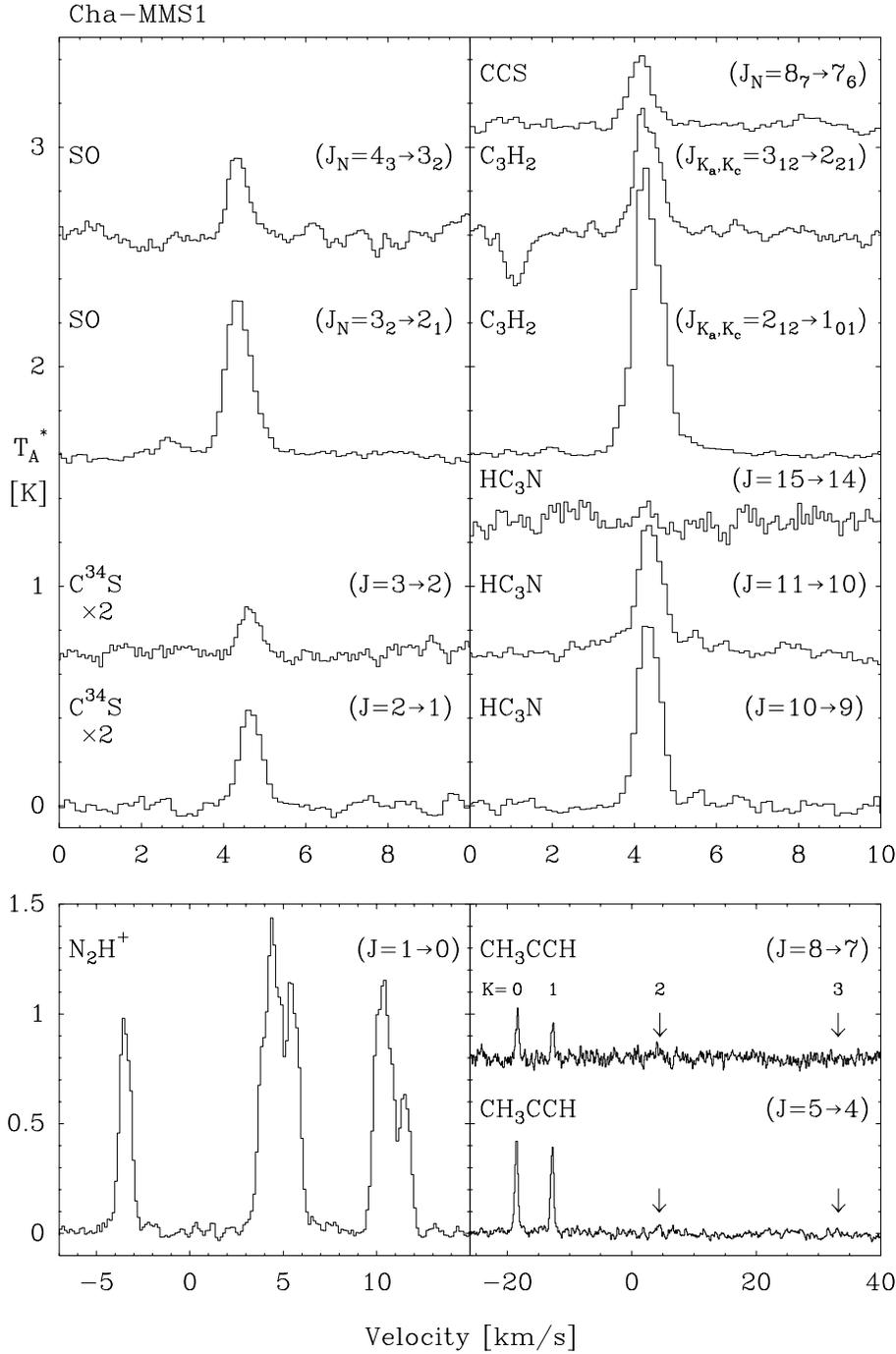}}
\caption[]{The observed spectra towards the centre of Cha-MMS1. The spectra are 
in the $T_{\rm A}^*$ scale and the velocity in relative to the local
standard of rest.  The upper panel shows the lines observed using
frequency switching, while the lower panel shows the position switched
spectra. Note the different velocity ranges.  The intensity in the C$^{34}$S spectra
are multiplied by a factor of two. The feature showing a negative intensity in the
C$_3$H$_2(J_{K_a,K_c}=3_{1,2}\rightarrow2_{2,1})$ spectrum is caused
by the $J_k=3_0\rightarrow2_0\, A^+$ line of CH$_3$OH, which lies
close in the frequency.  The frequency switching method causes the
line to appear with a negative intensity. The CH$_3$CCH spectra are
centered on the $K=2$ components, which, however, remain
undetected. The locations of the $K=0,1,2 \mbox{ and } 3$ components
are marked by arrows.}
\label{figure:chaspectra} 
\end{figure*}

\begin{figure*} 
\resizebox{12cm}{!}{\includegraphics{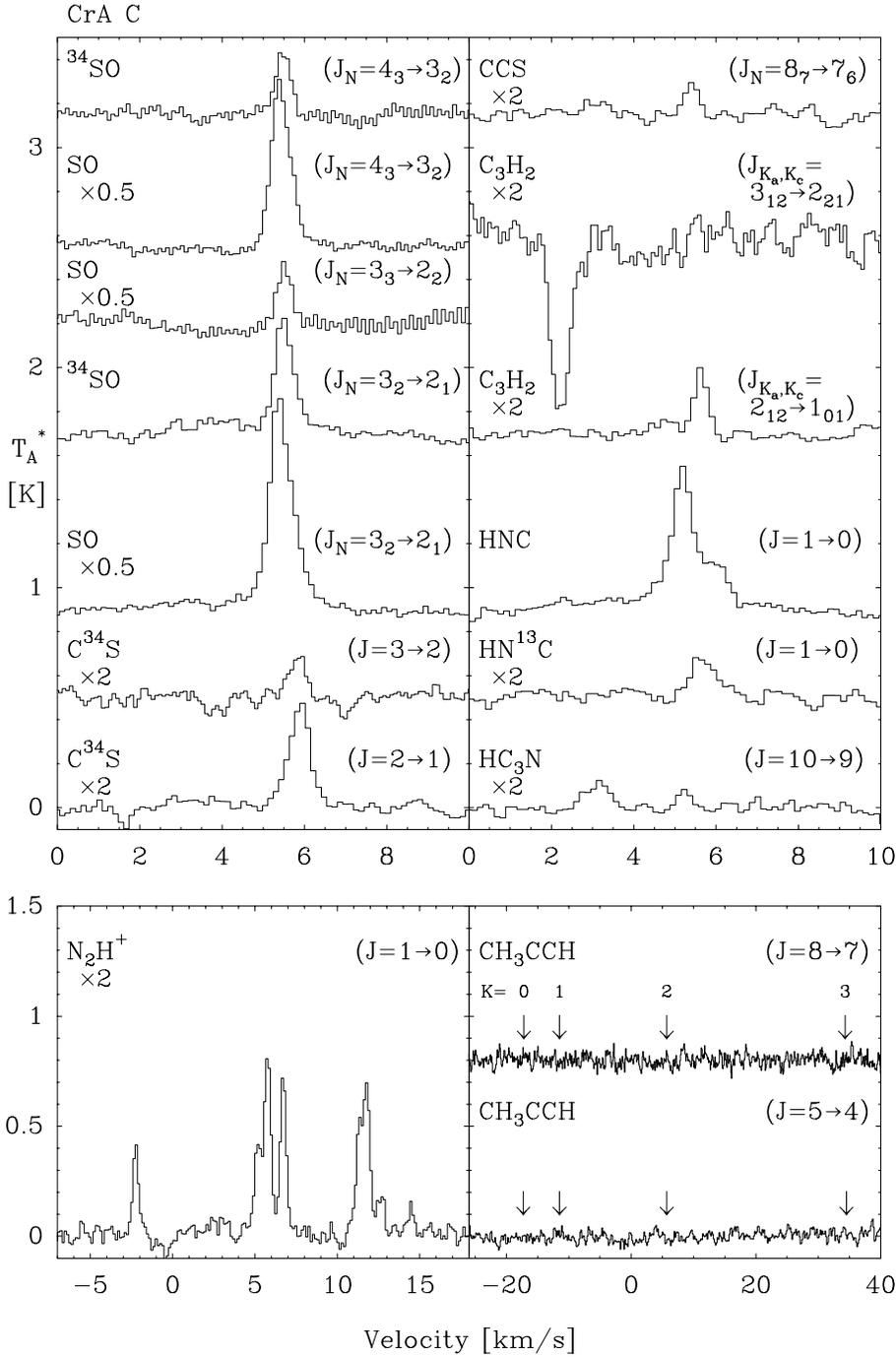}}
\caption[]{The observed spectra towards the centre of CrA~C. The spectra are 
in the $T_{\rm A}^*$ scale and the velocity in relative to the local
standard of rest. As for Cha-MMS1, the frequency switched and position
swithed spectra are placed into different boxes. The selection of
transitions is, however, slightly different. Note that many of the
spectra are either multiplied or divided by a factor of two.  The
C$_3$H$_2(J_{K_a,K_c}=3_{12}\rightarrow2_{21})$ spectrum contains the
negative `CH$_3$OH feature'.  In the CH$_3$CCH spectra we have marked
the positions where the $K=0,1,2 \mbox{ and } 3$ components should
lie.}
\label{figure:craspectra} 
\end{figure*}

One location in each core was observed in several molecular
transitions.  In Cha-MMS1 this position lies towards the HC$_3$N peak
at the offset $(-60\arcsec,-60\arcsec)$, $17\arcsec$ southeast of the
dust continuum peak.  In CrA~C the selected position lies close to the
SO peak at the offset $(30\arcsec,60\arcsec)$ from the map centre,
which has the highest SO line intensities. The spectra obtained
towards these two locations are shown in Figs.~\ref{figure:chaspectra}
and \ref{figure:craspectra}. \\

In Cha-MMS1 all observed transitions except
HC$_3$N($J=15\rightarrow14$) were detected.  The energy of the upper
state of this transition is rather high (52\,K) and so the
non-detection of this line in a cold cloud is unsurprising. Since
the SO lines were rather weak ($T_{\rm A}^* \leq 0.7$ K) and showed no
self-absorption features, $^{34}$SO was not observed. In CH$_3$CCH,
only the $K=0$ and $K=1$ components of the $J=5\rightarrow4$ and
$J=8\rightarrow7$ transitions were clearly detected.  The upper limits
(3$\sigma$) for the $K$=2 and $K$=3 components are $T_{\rm
A}^*$=0.063\,K ($J=5\rightarrow4$) and $T_{\rm A}^*$=0.045\,K
($J=8\rightarrow7$), respectively.  \\

In CrA~C, both of the observed SO transitions ($J_N=3_2\rightarrow2_1$
and $4_3\rightarrow3_2$) were bright ($T_{\rm A}^* > 1.5$~K). Only a
tentative detection of HC$_3$N($J=10\rightarrow9$) was made at a level
of $T_{\rm A}^*$=0.05 K, and so ther transitions were not
attempted. None of the CH$_3$CCH lines were detected, and the rest of the
lines, both from carbonaceous species as well as from N$_2$H$^+$,
were weaker than in Cha-MMS1. 

The common isotopomer HNC shows a broad, asymmetric profile.  The fact
that the peak is shifted with respect to the other lines, in
particular that of HN$^{13}$C, can be explained by absorption in the low
density halo around the core.  The larger width of this transition is
partly due to the hyperfine structure discussed in Sect.~2.1.
However, the broad wings visible on both sides of the line centre 
are not seen in any other line (including
C$^{18}$O). The presence of high velocity gas towards this location
should be tested in other optically thick transitions. 

In both clouds the C$_3$H$_2(J_{K_a,K_c}=3_{12}\rightarrow2_{21}$)
spectra also contain the $J_k=3_{-1}\rightarrow 2_{-1}\, E$ and
$J_k=3_{0}\rightarrow 2_{0}\, A^+$ lines of CH$_3$OH. The former
component appears in Figs.~\ref{figure:chaspectra} and
\ref{figure:craspectra} as a negative spike due to the frequency
switching procedure. \\

The observed line parameters (from fits of Gaussian line-profiles) in
the selected positions are given in Table~\ref{table:lineparameters}.
We list the peak antenna temperatures ($T_{\rm A}^*$), the radial
velocities ($v_{\rm LSR}$), the full-width at half-maximum (FWHM)
line-widths ($\Delta v$), and the velocity-integrated line-intensities
($\int T_{\rm A}^*\,dv$) of the detected lines for both cores. For the
undetected lines, the upper limits (3$\sigma$) for $T_{\rm A}^*$ are
given.

\subsection{Column densities}

The method used in the calculation of the excitation temperatures,
optical depths, and column densities is described in Appendix~A. The
results of this analysis are summarized In
Table~\ref{table:columndensities}.  The columns of this Table are: (1)
the molecule; (2) the transition; (3), (4) and (5) the excitation
temperature ($T_{\rm ex}$) and the line-centre optical depth
($\tau_{\nu_{ul}}$) of the transition, and the total column density
($N_{\rm tot}$) of the molecule in Cha-MMS1; (6), (7) and (8) $T_{\rm
ex}$, $\tau_{\nu_{ul}}$ and $N_{\rm tot}$ for the same molecule and
transition in CrA~C. \\

We have estimated the column densities of C$^{18}$O towards the
centres of CrA C and Cha-MMS1 using data from \cite{harju93} and
Haikala et al. 2000 (in preparation). Assuming optically thin line
emission and a uniform excitation of the energy levels with $T_{\rm
ex}=$ 10~K, we obtained $N({\rm C^{18}O})\approx 3\,10^{15}$ cm$^{-2}$
in CrA~C, and $N({\rm C^{18}O})\approx 2.5\,10^{15}$ cm$^{-2}$ in
Cha-MMS1. The two sources thus appear to have very similar C$^{18}$O
column densities. The column densities of the observed molecules
relative to C$^{18}$O are presented in Table~\ref{table:fractions}.
The conversion of the ratios listed in this Table to fractional
abundances relative to H$_2$ is not trivial. The fractional CO
abundance is expected and has also been observed to change from cloud
to cloud. As discussed by \cite{harjunpaa} the fractional C$^{18}$O
abundances can be different in star forming and non-star forming
regions. The relations they derived towards the R CrA and the Cha I
clouds (assuming $N({\rm H_2})/E(J-K) = 5.4
\, 10^{21}$ cm$^{-2}$ mag$^{-1}$) give $2.4 \, 10^{22}$ cm$^{-2}$ and $1.2
\, 10^{22}$ cm$^{-2}$ for $N({\rm H_2})$ in CrA C and Cha-MMS1,
respectively. However, neither region studied by \cite{harjunpaa} in
these complexes is actually close to our sources. The assumption that
the relation they derived towards Coalsack is valid in the quiescent
CrA C would result in an H$_2$ column density of $3.8 \, 10^{22}$
cm$^{-2}$. In comparison, the relation
between $N({\rm H_2})$ and $N({\rm C^{18}O})$ derived by
\cite{frerking} in Taurus yields $N({\rm H_2}) = 1.9
\, 10^{22}$ cm$^{-2}$ and $N({\rm H_2}) = 1.6 \, 10^{22}$ cm$^{-2}$ towards
CrA C and Cha-MMS1, respectively. All these H$_2$ column densities 
are similar within a factor of two for each source, and so 
the values presented in Table~\ref{table:fractions} may be
converted to approximate fractional abundances by multiplying them by
$2\times 10^{-7}$, bearing in mind that for CrA C, the value $1 \times
10^{-7}$ is possibly more appropriate.

Table~\ref{table:fractions} shows that the column
densities of carbon chain molecules (CCS, HC$_3$N, CH$_3$CCH and
C$_3$H$_2$) relative to C$^{18}$O are much larger towards Cha-MMS1 than
towards CrA C, for which upper limits only could be derived for all
species other than CCS. The HC$_3$N/C$^{18}$O column density ratio towards Cha-MMS1 
is almost 0.2, which implies that the fractional HC$_3$N abundance is of order
$10^{-8}$, being several times larger than towards
the cyanopolyyne maximum in TMC-1 (\cite{pratap}).  The SO/C$^{18}$O
abundance ratio is larger towards CrA C. Inspection of the intensity
ratios of the $^{34}$SO and the normal isotope SO lines reveals that the latter
are optically thick or even self-absorbed towards CrA C, and the
derived SO column densities there are almost certainly lower limits
(the terrestial $^{32}$S/$^{34}$S isotope ratio is 22.5 and in the
local interstellar medium values larger than 10 are normally assumed).
On the other hand, the N$_2$H$^+$/C$^{18}$O column density ratio is a
few times larger towards Cha-MMS1 than CrA C.  Two molecules
which have similar column densities towards both cores are C$^{34}$S and
CH$_3$OH. For the rest of the molecules, the column densities relative 
to C$^{18}$O are so different in the two cores that despite the uncertainty 
concerning the $N({\rm H_2})/N({\rm C^{18}O})$ conversion factor, this
issue may be discussed below in terms of differences in fractional abundances.  

Chemical differences between the two cores are manifest in 
Table~\ref{table:ratios}, where we list the column density ratios of
several molecules. The difference is most marked in the HC$_3$N/SO
column density ratio, which is at least three orders of magnitude
higher towards Cha-MMS1 than CrA C. The ratios HC$_3$N/CCS
and HC$_3$N/N$_2$H$^+$ are more than a hundred times larger, and the
C$^{34}$S/$^{34}$SO ratio is about 30 times larger towards Cha-MMS1
than towards Cra C. However, the CCS/N$_2$H$^+$ column density ratio
is similar towards both cores.

\section{Discussion}

In the previous Section it became evident that carbon-chain molecules
are considerably more abundant in Cha-MMS1 than in CrA C, whereas the
opposite is true for SO. In particular, the HC$_3$N/SO column density
ratio is at least 2000 times higher in Cha-MMS1 than in CrA~C. The
fact that HC$_3$N, CH$_3$CCH, c-C$_3$H$_2$ and CCS were detected
without difficulty in one of the cores, but were all very weak or
undetected in the other, is in agreement with the common belief that
these species thrive in the same regions. For CCS, the difference in
the abundances towards the two cores is the smallest for carbon-chain
molecules, being `only' about one order of magnitude.

As discussed in Sect.~2.2, time-dependent models of gas-phase
chemistry suggest that HC$_3$N and other carbon-chain molecules reach
their peak abundances at early stages of chemical evolution, whereas
SO becomes abundant first at later times. Adopting this view, the
HC$_3$N/SO abundance ratios would indicate that Cha-MMS1 is chemically
younger than CrA C. In addition, the CS/SO abundance ratio has been
suggested to depend on the cloud age (\cite{bergin97}; \cite{ruffle99}; 
\cite{nilsson}). Indeed, the CS/SO ratio is clearly larger towards
Cha-MMS1, which is mainly due to variation of the SO abundance, since
the CS abundance is rather similar towards both cores.

On the other hand, it is difficult to understand in terms of pure
gas-phase chemistry why CrA C has a higher SO abundance but a lower
N$_2$H$^+$ abundance than Cha-MMS1: The production of N$_2$H$^+$ also
increases first at later stages of chemical evolution. In particular,
the fact that CCS/N$_2$H$^+$ column density ratio is similar towards
both cores seems to contradict the expectation presented in Sect.~2.2
that this ratio could be usable as a chemical clock.

The inclusion of freezing-out of atoms, ions and molecules onto grain
surfaces causes a complication for the time-dependence of molecular
abundances. According to \cite{ruffle97}, a second cyanopolyyne peak
may occur when the depletion of gas-phase species becomes
significant. The SO abundance also increases considerably at this time
(\cite{bergin97}; \cite{ruffle99}). In the models of \cite{ruffle99},
the SO peak comes slightly {\sl after} the second HC$_3$N peak (see their
Figs. 1 and 2), {probably because the C$^+$ abundance is then
lower}. After the SO peak, the dominant species in the gas phase are
CO, CS and SO, until they too are depleted. It is uncertain how
comparable the models of \cite{bergin97} and \cite{ruffle99} are, however, 
since the former do not predict any late cyanopolyyne maximum.

N$_2$H$^+$ is included in the models of \cite{rawlings} and \cite{bergin97},
where depletion onto grains are taken into account. In both models
N$_2$H$^+$ shows a late peaking, and two reasons for this peaking are
given. First, N$_2$, which contributes to the N$_2$H$^+$
production, is only loosely bound in grain surfaces and returns quickly to
the gas phase. Second, the drop in the gas-phase abundances {of CO, 
H$_2$O and other neutral molecules} with the advance of freezing-out decreases 
the destruction rate of N$_2$H$^+$, until the electron recombination starts 
to dominate its removal. 

Finally, it should be noted that nearby or embedded young stars may
revive the ion-molecule chemistry in a dense core.  As discussed by
\cite{hartquist96}, an intensified  desorption of molecules from grain 
surfaces, the subsequent fragmentation of molecules into
atoms in the gas-phase, and finally the ionization of atoms in a
star-forming core may lead to increased abundances of C and C$^+$.
These could then re-start the synthesis of complex molecules, and
accordingly the chemistry would have an `early-time' character.

We have seen that chemistry models, regardless of whether they take
accretion onto grains into account or not, predict that SO and
N$_2$H$^+$ peak at late stages of chemical evolution, and that SO
peaks later than HC$_3$N during the pre-stellar phase.  Therefore,
concerning the cores studied here, it is difficult to avoid the
following conclusions: 1) CrA~C has reached an advanced chemical
stage.  2) Cha-MMS1 either represents the so called late-time
cyanopolyyne peak or a later stage where the chemistry has been
influenced by neighbouring young stars.

The latter conclusion is based on the large N$_2$H$^+$ abundance in
Cha-MMS1. Moreover, the interpretation that a collapsing cloud core
containing a protostar, as Cha-MMS1 probably does, has reached the second,
`late-time' cyanopolyyne peak seems more natural than that it would
represent `early-time' chemistry, since such a core should exhibit a high
degree of depletion (\cite{rawlings}; \cite{bergin97}).
Possible outflows and radiation from the embedded protostar, IRS~10,
or the other nearby protostars may have influenced the chemical
composition of Cha-MMS1.  \cite{suzuki} {found that} NH$_3$ tends to be 
abundant in star-forming regions, and it seems possible that the production 
of NH$_3$, and therefore also that of N$_2$H$^+$ (see Sect.~2), somehow depend 
on desorption processes. On the other hand, little time has passed since the collapse
began (which should be coincident with the second cyanopolyyne peak) 
because the duration of the Class 0 phase is very  short ($\sim 10^4$ years),  
and it is questionable whether the chemistry has really changed significantly 
during this time. 

According to the models of \cite{ruffle99} and \cite{bergin97}, where the
chemical development is coupled to {kinematical} models, the
compositions observed in Cha-MMS1 and CrA C correspond to times a few
million years after the initial state, which in these studies has been
defined as a state where hydrogen is molecular but all other species are
found as atoms or atomic ions.  If the physical characteristics of the two
cores were similar, Cha-MMS1 would not need to be much younger than CrA C.
The latter core, however, appears to have a lower average density, 
and therefore the chemical evolution has probably been slower there. The result is
surprising, since star-forming cores of the type Cha-MMS1 are normally
expected to be older than starless cores like CrA~C. Furthermore, previous
observations have shown that the carbon-chain molecules are less abundant in
star-forming regions than in quiescent clouds (\cite{suzuki}).

CrA~C has been estimated to be gravitationally bound and therefore a
potential site of star-formation (\cite{harju93}). The present results
indicate, however, that CrA C has existed long enough to reach an advanced
chemical stage, which according to chemistry models implies a minimum age of
$\sim10^6$ years. This longevity raises doubts about the star-formation
capacity of the core.

In quiescent cores, star formation occurs due to the cloud's
self-gravity, which has to work against the thermal pressure and the
pressure and tension of the magnetic fields. Line emission from
molecules, especially CO (as long as it is abundant in the gas phase),
{and collisions between gas and dust particles}, can keep a cloud
nearly isothermal even when contracting, and the importance of the
magnetic support in weakly ionized cores has been recognized as the
dominant factor preventing the collapse within the free-fall time,
which in the case of a dense core is of the order of $10^5$ years.
The magnetic support eventually gives way in the {central parts of
a} core due to ambipolar diffusion, resulting in gravitational
instability and collapse (\cite{spitzer}; \cite{mouschovias79};
\cite{lizano}). {The density enhancement required for the gravitational 
collapse depends on the initial mass-to-magnetic-flux ratio. The
rate at which the central density increases is related to the
timescale of ambipolar diffusion (see e.g. Sect. 2 in
\cite{ciolek93}).}  The timescale of ambipolar diffusion in a dense
core is proportional to the fractional ionization, and is typically of
the order of $10^6$ years (e.g. \cite{spitzer}; \cite{hartquist89}).
This time is shorter than the time needed for reaching a chemical
equilibrium in pure gas-phase models, but comparable with the
depletion timescale in models in which accretion onto grain surfaces
are taken into account. According to \cite{hartquist89}, depletion can
influence the dynamical state of a core, and in the extreme case, even
prevent it from collapsing by causing an increase in the fractional
ionization and removing coolants. {Furthermore, as pointed out by
Ciolek \& Mouschovias (1993, 1994,1995), the properties of dust grains
can affect the ambipolar diffusion timescale and thus
regulate the evolution of a core.}

In simulations by \cite{vanhala}, the state of a gas-cloud first
becomes non-adiabatic (i.e. the cloud is allowed to collapse) when its
temperature has risen to about 27\,K, at which point the rotational
levels of H$_2$ become excited. Even though these authors use perhaps
unrealistically low CO cooling efficiences (they neglect the cooling
below 10 K) this model may be qualitatively valid for very cold cores
with highly reduced abundances of heavier molecules. If the
temperature is too low, then the excitation is insufficient to provide
strong enough line-emission, and a thermal balance is
established. Hence, it may be that such cores have a poorer
star-formation ability than previously supposed.  Furthermore, an
elevated temperature possibly is a useful diagnostic of the onset of a
collapse, and could be used to identify dense pre-stellar cores that are
close to forming stars.

{Inferring from its chemical composition,} the development in
CrA C seem to have been very slow indeed, or then the gravitational
forces in the core are simply balanced by thermal and magnetic
pressures. {In the ambipolar diffusion model the slow dynamical
evolution could be explained by 1) a low initial mass-to-magnetic-flux
ratio in the surrounding cloud; or 2) a high degree of ionization or a
high fraction of charged dust grains, which both would lengthen the
ambipolar diffusion timescale.}  The fact that the N$_2$H$^+$
abundance is lower in CrA C than in Cha-MMS1 suggests that the
fractional ionization is indeed higher in the former.  An alternative
reason for the slow progress is that the cooling rate is very
low in this core. Independent determinations of the kinetic
temperature, the degree of ionization, and the degree of depletion
towards this core would certainly clarify this issue. Based on the
results of coupled chemical and {kinematical} models
(e.g. \cite{rawlings}; \cite{bergin97}), the high degree of depletion
inferred from the molecular abundances implies a very advanced stage
of evolution, and if the core is going to collapse, then this has
already begun, or will do so soon deep in its interior parts. A search
for infalling motions towards CrA C could be therefore profitable.

{According to} \cite{lehtinen}, the region of Cha-MMS1 may be an example
of sequential star-formation, in which the core collapse is assisted
by external forces such as stellar-winds originating from nearby young
stars. {This would conform with a rapid development, and may reflect
favourable initial conditions for contraction and gravitational
collapse in the parental cloud.}  In order to judge whether the
chemistry in Cha-MMS1 is controlled by desorption or depletion, it
would be useful to observe molecules which are predicted to increase
in abundance when desorption mechanisms are effective.  Our tentative
column density estimates of CH$_3$OH (an example of such a molecule)
did not, however, show any difference in relative abundances between
CrA C and Cha-MMS1. Detailed chemistry models of the relative timing
of the HC$_3$N, SO and N$_2$H$^+$ peaks would therefore be valuable.

The tiny SO maxima lying symmetrically on both sides of the HC$_3$N
core in Cha-MMS1 (see Fig.~\ref{figure:chamaps}) may be related to an
outflow from the embedded protostar. These emission peaks and the
Herbig-Haro object HH49/50 lie on the same line intersecting the
centre of Cha-MMS1, and it seems that Cha-MMS1 can indeed be the
driving source of HH49/50, as suggested by \cite{reipurth}. The
stronger SO line emission towards the outflow axis may be understood
in terms of an increased efficiency of neutral-neutral reactions
involved in the formation of SO in shock-heated gas.

\section{Conclusions}

Comparison of observed molecular column densities and abundances in
two dense cores, Cha-MMS1 and CrA~C, reveal that they have very
different chemical compositions, which according to time-dependent
chemistry models can be interpreted as representing different stages
of chemical evolution.  Cha-MMS1, which is dynamically more evolved
and contains a Class 0 protostar (\cite{reipurth}; \cite{lehtinen}),
has large abundances of carbon-chain molecules. This is normally
interpreted as a signature of an early stage of chemical evolution. In
contrast, CrA~C, which we suspected to be virginal and therefore a
likely fund of `early type' molecules, has, to all appearances,
reached chemical maturity.  This result seems to contradict the common
assumption that star-forming cores are (chemically) older than
starless cores. The chemical difference between the two cores is
particularly pronounced in the HC$_3$N/SO abundance ratio, which is
about 2000 times higher in Cha-MMS1 than in CrA~C.

The fact that Cha-MMS1 has a large N$_2$H$^+$ abundance arouses doubts
about its chemical youth, however. Namely, this molecular ion is
formed from N$_2$, which should become abundant only at later times.
In fact, Cha-MMS1 probably represents the `late-time' cyanopolyyne
peak predicted by \cite{ruffle97}, which occurs when the freezing-out
of molecules onto grain surfaces begins to be
significant. Alternatively, the characteristics of youthful chemistry
in Cha-MMS1 could have developed due to outflows and ionizing
radiation from the embedded protostar or the neighbouring newly born
stars. It should be noted, however, that these two alternatives do not
have to lie very much apart in time. A high degree of depletion should
be coincident with the collapse of the core nucleus
(e.g. \cite{rawlings}; \cite{bergin97}), and Cha-MMS1 still contains a
protostar in its main accretion phase.  The fact that Cha-MMS1 is
associated with a small cluster of young stars indicates a high
star-formation efficiency in this region (\cite{lehtinen}), which
would be consistent with a rapid collapse.

We have detected distinct SO emission maxima on both sides of Cha-MMS1,
which may indicate the presence of a north-south oriented bipolar
outflow.  The suggestion by \cite{reipurth} that Cha-MMS1 contains the
central source of the Herbig-Haro object HH49/50 is strongly supported
by the fact the SO peaks, Cha-MMS1 and HH49/50 lie along the same line.

The situation observed in CrA C, i.e. low abundances of carbon chain
molecules and the large SO and N$_2$H$^+$ abundances, correspond to
final stages of pure gas-phase chemistry evolution or to the `SO peak'
which comes after the late cyanopolyyne peak in the model of
\cite{ruffle99}, in which gas-grain interactions has been taken into
account.  The lack of star-formation activity in the chemically `old'
core CrA~C shows that the dynamical evolution has been slow there or
that the core may have even reached an equilibrium. This may be due to
{a low mass-to-magnetic-flux ratio in the surrounding cloud, or 
a high degree of ionization or a high fraction of charged dust grains 
making the ambipolar diffusion time very long (e.g. \cite{ciolek94})}. 
On the other hand, insufficient cooling efficiency due to freezing-out of 
heavy molecules may have increased the thermal support.

\begin{acknowledgements}

We thank Lic. P\"aivi Harjunp\"a\"a for discussions concerning the
`LTE' method applied here. We are grateful to Dr. Mark Rawlings,
Dr. Malcolm Walmsley, Dr. Paola Caselli, and the referee, Dr. Glenn
Ciolek, for very helpful comments on the manuscript. The work by S.K.,
J.H. and A.H. was supported by the Academy of Finland through grant
No. 1011055. The Swedish-ESO Submillimetre Telescope is operated
jointly by ESO and the Swedish National Facility for Radio Astronomy,
Onsala Space Observatory at Chalmers University of Technology.

\end{acknowledgements}

\clearpage
\begin{table*}
\caption[Molecules]{Spectroscopic properties of the observed molecules.} 
\begin{tabular}{llccrc} \hline 

Molecule  & Type  & $\mu_{\rm electric}$ &  $A$ &\multicolumn{1}{c}{$B$}  & $C$ \\
          &       & [Debye] & [MHz]&\multicolumn{1}{c}{[MHz]}& [MHz] \\ \hline
SO        &linear ($^3\Sigma$)    &1.55$\;\:$&--                      &21\,523.020$\;\:$      &--\\
$^{34}$SO &linear ($^3\Sigma$)    &1.55$\;\:$&--                      &21\,102.720$\;\:$      &--\\
C$^{34}$S &linear ($^1\Sigma$)    &1.957     &--                      &24\,103.541$\;\:$      &--\\
c-C$_3$H$_2$&asymmetric top (o/p) &3.43$\;\:$&35\,092.6$\;\:\;\:$     &32\,212.8$\;\:\;\:\;\:$&16\,749.1$\;\:\;\:\;\:$\\
CH$_3$OH  &asymmetric top ($A$)   &0.896     &127\,628.193$\;\:$      &24\,687.552$\;\:$      &23\,754.594$\;\:$ \\
CH$_3$OH  &asymmetric top ($E$)   &0.896     &127\,628.7051           &24\,693.0658           &23\,757.1371 \\
CCS       &linear ($^3\Sigma$)    &2.81$\;\:$&--                      &6\,477.750$\;\:$       &--\\
HC$_3$N   &linear ($^1\Sigma$,hfs)&3.724     &--                      &4\,549.058$\;\:$       &--\\
CH$_3$CCH &symmetric top ($A/E$)  &0.75$\;\:$&158\,590.0$\;\:\;\:\;\:$&8\,545.86$\;\:\;\:$    &8\,545.86$\;\:$\\
HNC       &linear ($^1\Sigma$,hfs)&3.05$\;\:$&--                      &45\,331.990$\;\:$      &--\\
HN$^{13}$C&linear ($^1\Sigma$,hfs)&2.699     &--                      &43\,545.61$\;\:\;\:$   &--\\
N$_2$H$^+$&linear ($^1\Sigma$,hfs)&3.40$\;\:$&--                      &46\,586.867$\;\:$      &--\\
\hline
\end{tabular}
\label{table:molecules}
\end{table*}

\begin{table*}
\caption[]{Map centre coordinates and the positions selected for pointed
observations.} 
\begin{tabular}{lccrr}\hline
\multicolumn{1}{c}{Core} & \multicolumn{2}{c}{Map centre} & 
\multicolumn{2}{c}{Peak position} \\                
 & $\alpha$ & $\delta$ &  
   $\Delta\alpha$ & $\Delta\delta$ \\
 & \multicolumn{2}{c}{(1950.0)} & &  \\ \hline

Cha-MMS1 & $11^{\rm h}05^{\rm m}28\fs0$ & $-77\degr06\arcmin32\arcsec$ &  
$-60\arcsec$ & $-60\arcsec$ \\

CrA~C    & $19^{\rm h}00^{\rm m}33\fs2$ & $-37\degr20\arcmin22\arcsec$ &   
$30\arcsec$ &  $60\arcsec$ \\ \hline             
\end{tabular}
\label{table:coordinates}
\end{table*}

\begin{table*}
%\caption[Molecules]{Observed transitions arranged according to frequency,
%line strength, energy of the upper state, \\
%and some telescope parameters at the corresponding frequency.}
\caption[Molecules]{Observed transitions and some antenna parameters
at their frequencies.}
\begin{tabular}{rlcll|cc} \hline

\multicolumn{1}{c}{Frequency}&Molecule&Transition&$S_{\rm ul}$&
\multicolumn{1}{c|}{$E_{\rm u}/{\rm k}$} & $\Theta_{\rm mb}$ & $\eta_{\rm mb}$ \\

\multicolumn{1}{c}{[MHz]}&&&&\multicolumn{1}{c|}{[K]} &
[$\arcsec$] &  \\ \hline

85\,338.893 &ortho-c-C$_3$H$_2$&$J_{K_{\rm a},K_{\rm c}}=2_{1,2} \rightarrow1_{0,1}$&$\;\:$1.500          &$\;\:$4.095$^{\rm a}$  & 59 & 0.75 \\
85\,457.271 & CH$_3$CCH        &$J_K=5_0 \rightarrow 4_0\, A$                       &$\;\:$5.000          &12.304           & 59 & 0.75 \\
85\,455.622 & CH$_3$CCH        &$J_K=5_1 \rightarrow 4_1\, E$                       &$\;\:$4.800          &11.484$^{\rm b}$ & 59 & 0.75 \\
87\,090.859 &HN$^{13}$C        &$J(F)=1(2) \rightarrow 0(1) $                       &$\;\:$1.000$^{\rm c}$&$\;\:$4.180  & 58 & 0.75 \\
90\,663.543 &HNC               &$J(F)=1(2) \rightarrow 0(1) $                       &$\;\:$1.000$^{\rm c}$&$\;\:$4.351  & 56 & 0.74 \\
90\,978.993 &HC$_3$N           &$J=10 \rightarrow 9 $                               &10.000               &24.015           & 55 & 0.74 \\
93\,173.809 &N$_2$H$^+$        &$J(F_1,F)=1(2,3) \rightarrow 0(1,2)$                &$\;\:$1.000$^{\rm c}$&$\;\:$4.472  & 54 & 0.74 \\
93\,870.023 &CCS               &$J_N = 8_7 \rightarrow 7_6$                         &$\;\:$7.971          &19.892           & 54 & 0.74 \\
96\,412.940 &C$^{34}$S         &$J=2 \rightarrow 1$                                 &$\;\:$2.000          &$\;\:$6.941            & 52 & 0.73 \\
97\,715.388 &$^{34}$SO         &$J_N=3_2 \rightarrow 2_1$                           &$\;\:$2.935          &$\;\:$9.100            & 52 & 0.73 \\
99\,299.879 &SO                &$J_N=3_2 \rightarrow 2_1$                           &$\;\:$2.933          &$\;\:$9.226            & 51 & 0.73 \\
100\,076.389&HC$_3$N           &$J=11 \rightarrow 10$                               &11.000               &28.818           & 50 & 0.73 \\
129\,138.898&SO                &$J_N=3_3 \rightarrow 2_2$                           &$\;\:$2.670          &25.511           & 39 & 0.68 \\
135\,775.633&$^{34}$SO         &$J_N=4_3 \rightarrow 3_2$                           &$\;\:$3.935          &15.609           & 37 & 0.67 \\
136\,464.400&HC$_3$N           &$J=15 \rightarrow 14$                               &15.000               &52.395           & 37 & 0.67 \\
136\,728.010&CH$_3$CCH         &$J_K=8_0 \rightarrow 7_0\, A$                       &$\;\:$8.000          &29.529           & 37 & 0.67 \\
136\,725.397&CH$_3$CCH         &$J_K=8_1 \rightarrow 7_1\, E$                       &$\;\:$7.975          &28.708$^{\rm b}$ & 37 & 0.67 \\
138\,178.648&SO                &$J_N=4_3 \rightarrow 3_2$                           &$\;\:$3.938          &15.857           & 36 & 0.67  \\
144\,617.109&C$^{34}$S         &$J=3 \rightarrow 2$                                 &$\;\:$3.000          &13.881           & 35 & 0.66 \\
145\,089.595&ortho-c-C$_3$H$_2$&$J_{K_{\rm a},K_{\rm c}}=3_{1,2} \rightarrow2_{2,1}$&$\;\:$1.250          &13.699$^{\rm a}$ & 35 & 0.66  \\
145\,097.470&CH$_3$OH          &$J_k=3_{-1} \rightarrow 2_{-1}\, E$                 &$\;\:$2.667          &11.607$^{\rm b}$ & 35 & 0.66 \\
145\,103.230&CH$_3$OH          &$J_k=3_{0} \rightarrow 2_{0}\, A^+$                 &$\;\:$3.000          &13.929           & 35 & 0.66 \\
\hline
\multicolumn{7}{l}{$^{\rm a}$ Relative to the ortho ground-state.} \\
\multicolumn{7}{l}{$^{\rm b}$ Relative to the $E$-symmetry species
ground-state.} \\
\multicolumn{7}{l}{$^{\rm c}$ This value refers to the $J=1\rightarrow 0$
transition when all hfs components have been accounted for.}
\end{tabular}
\label{table:lines}
\end{table*}

\clearpage

\begin{table*}
\caption[Parameters]{Observational line parameters towards the centres of
Cha-MMS1 and CrA~C. \\
The transitions are listed in the same order as in the figures showing the
spectra.}
\fontsize{8}{11}
\selectfont
\begin{tabular}{ll|cccc|cccc} \hline

Molecule & Transition & \multicolumn{4}{|c|}{Cha-MMS1} & 
\multicolumn{4}{|c}{CrA~C} \\
  & &\multicolumn{1}{c}{$T_{\rm A}^*$} &\multicolumn{1}{c}{$v_{\rm
     LSR}$} &\multicolumn{1}{c}{$\Delta v$} &\multicolumn{1}{c|}{$\int
     T_A^* {\rm d} v$} &\multicolumn{1}{c}{$T_{\rm A}^*$}
     &\multicolumn{1}{c}{$v_{\rm LSR}$} &\multicolumn{1}{c}{$\Delta
     v$} &\multicolumn{1}{c}{$\int T_A^* {\rm d} v$} \\ &
     &\multicolumn{1}{c}{[K]} &\multicolumn{1}{c}{[km\,s$^{-1}$]}
     &\multicolumn{1}{c}{[km\,s$^{-1}$]}
     &\multicolumn{1}{c|}{[K\,km\,s$^{-1}$]} &\multicolumn{1}{c}{[K]}
     &\multicolumn{1}{c}{[km\,s$^{-1}$]}
     &\multicolumn{1}{c}{[km\,s$^{-1}$]}
     &\multicolumn{1}{c}{[K\,km\,s$^{-1}$]} \\ \hline 

$^{34}$SO &$4_3 \rightarrow 3_2$ & -& - & - & -
&$0.29\!\pm\!0.02$&$5.48\!\pm\!0.01$&$0.43\!\pm\!0.03$&$0.13\!\pm\!0.02$\\

SO &$4_3 \rightarrow 3_2$ &$0.37\!\pm\!0.04$&$4.35\!\pm\!0.02$
&$0.55\!\pm\!0.04$&$0.21\!\pm\!0.04$&$1.44\!\pm\!0.04$&$5.43\!\pm\!0.01$&
$0.56\!\pm\!0.01$&$0.87\!\pm\!0.04$\\

SO &$3_3 \rightarrow 2_2$ & -& - & - & -
  &$0.53\!\pm\!0.06$&$5.52\!\pm\!0.02$&$0.39\!\pm\!0.04$&$0.20\!\pm\!0.06$\\

$^{34}$SO     &$3_2 \rightarrow 2_1$          & -& -  & -  & -
     &$0.50\!\pm\!0.03$&$5.51\!\pm\!0.01$&$0.57\!\pm\!0.03$&$0.33\!\pm\!0.04$\\

SO            &$3_2 \rightarrow 2_1$
     &$0.70\!\pm\!0.02$&$4.37\!\pm\!0.01$
&$0.74\!\pm\!0.02$&$0.56\!\pm\!0.03$&$1.79\!\pm\!0.05$&$5.43\!\pm\!0.01$
&$0.70\!\pm\!0.02$&$1.40\!\pm\!0.06$\\

C$^{34}$S     &$3 \rightarrow 2$              &$0.10\!\pm\!0.01$
&$4.64\!\pm\!0.02$&$0.53\!\pm\!0.04$&$0.05\!\pm\!0.01$&$0.09\!\pm\!0.02$
&$5.83\!\pm\!0.02$&$0.48\!\pm\!0.05$&$0.05\!\pm\!0.01$\\

C$^{34}$S     &$2 \rightarrow 1$              &$0.22\!\pm\!0.01$
&$4.66\!\pm\!0.01$&$0.63\!\pm\!0.03$&$0.14\!\pm\!0.01$
&$0.23\!\pm\!0.01$&$5.89\!\pm\!0.01$&$0.62\!\pm\!0.03$&$0.16\!\pm\!0.02$\\

CCS           &$8_7 \rightarrow 7_6$          &$0.31\!\pm\!0.02$
&$4.15\!\pm\!0.01$&$0.65\!\pm\!0.03$&$0.21\!\pm\!0.02$&$0.07\!\pm\!0.01$
&$5.40\!\pm\!0.03$&$0.38\!\pm\!0.06$&$0.03\!\pm\!0.02$\\

o-c-C$_3$H$_2$&$3_{1,2} \rightarrow 2_{2,1}$ &$0.54\!\pm\!0.06$
&$4.31\!\pm\!0.02$&$0.73\!\pm\!0.04$&$0.43\!\pm\!0.05$&$<$0.26$^{\rm a}$& - & - &-\\

o-c-C$_3$H$_2$&$2_{1,2} \rightarrow 1_{0,1}$  &$1.24\!\pm\!0.02$
&$4.32\!\pm\!0.01$&$0.82\!\pm\!0.01$&$1.09\!\pm\!0.02$&$0.15\!\pm\!0.01$
&$5.63\!\pm\!0.01$&$0.38\!\pm\!0.03$&$0.07\!\pm\!0.01$\\

CH$_3$OH &$3_{-1} \rightarrow 2_{-1}\,
E$&$0.42\!\pm\!0.03$&$4.46\!\pm\!0.01$
&$0.62\!\pm\!0.02$&$0.28\!\pm\!0.03$&$0.61\!\pm\!0.03$&$5.63\!\pm\!0.01$
&$0.53\!\pm\!0.01$&$0.34\!\pm\!0.03$\\

CH$_3$OH &$3_{0} \rightarrow 2_{0}\,A^+$
&$0.48\!\pm\!0.03$&$4.46\!\pm\!0.01$&$0.62\!\pm\!0.02$&$0.32\!\pm\!0.03$
&$0.79\!\pm\!0.03$&$5.63\!\pm\!0.01$&$0.53\!\pm\!0.01$&$0.44\!\pm\!0.03$\\

HN$^{13}$C &$1 \rightarrow 0$ & -& - & - & -
&$0.09\!\pm\!0.01$&$5.70\!\pm\!0.03$&$0.76\!\pm\!0.08$&$0.07\!\pm\!0.02$\\

HNC &$1 \rightarrow 0$ & -& - & - & - &$0.63\!\!\pm\!\!0.05\!^{\rm
b}$&$5.21\!\!\pm\!\!0.02\!^{\rm b}$&$0.65\!\!\pm\!\!0.01\!^{\rm
b}$&$0.57\!\pm\!0.06$\\

HC$_3$N &$15 \rightarrow 14$ &$<0.13^{\rm a}$& - &- & - & - & - & - & - \\

HC$_3$N &$11 \rightarrow 10$
&$0.57\!\pm\!0.03$&$4.39\!\pm\!0.01$&$0.74\!\pm\!0.03$&$0.47\!\pm\!0.04$&
- & - & - & -\\

HC$_3$N &$10 \rightarrow 9 $
&$0.85\!\pm\!0.02$&$4.36\!\pm\!0.01$&$0.65\!\pm\!0.01$&$0.59\!\pm\!0.03$&
$0.04\!\pm\!0.01$&$5.23\!\pm\!0.04$&$0.34\!\pm\!0.09$&$0.02\!\pm\!0.01$\\

CH$_3$CCH &$8_0 \rightarrow 7_0$
&$0.20\!\pm\!0.02$&$4.53\!\pm\!0.02$&$0.58\!\pm\!0.04$&
$0.12\!\pm\!0.02$&$<$0.083$^{\rm a}$&-&-&-\\

CH$_3$CCH &$8_1 \rightarrow 7_1$
&$0.16\!\pm\!0.02$&$4.49\!\pm\!0.02$&$0.52\!\pm\!0.04$&
$0.08\!\pm\!0.02$&$<$0.083$^{\rm a}$&-&-&-\\

CH$_3$CCH &$5_0 \rightarrow 4_0$
&$0.43\!\pm\!0.01$&$4.41\!\pm\!0.01$&$0.63\!\pm\!0.02$&
$0.30\!\pm\!0.03$&$<$0.071$^{\rm a}$&-&-&-\\

CH$_3$CCH     &$5_1 \rightarrow 4_1$         &$0.39\!\pm\!0.01$&
$4.41\!\pm\!0.01$&$0.62\!\pm\!0.02$&$0.26\!\pm\!0.02$&$<$0.071$^{\rm a}$&-&-&-\\

N$_2$H$^+$ &$1\rightarrow 0$ & $1.67\!\pm\!0.03$&
$4.57\!\pm\!0.01$&$0.63\!\pm\!0.01$&$5.02\!\pm\!0.03$&$0.45\!\pm\!0.02$&
$5.77\!\pm\!0.01$&$0.46\!\pm\!0.01$&$0.94\!\pm\!0.03$\\
\hline
\multicolumn{10}{l}{$^{\rm a}\,3\sigma$.} \\
\multicolumn{10}{l}{$^{\rm b}$\, Applies for the strongest component in the
spectrum.} \\
\end{tabular}
\label{table:lineparameters}
\end{table*}

\begin{table*}
\caption[Columns]{The excitation temperatures and the line-centre optical
depths of the observed transitions, and the total \\ molecular column
densities towards the peak
positions of Cha-MMS1 and CrA~C.}
\fontsize{8}{11}
\selectfont
\begin{tabular}{ll|ccc|ccc} \hline
Molecule & Transition & \multicolumn{3}{|c|}{Cha-MMS1} & \multicolumn{3}{|c}{CrA~C}
\\
         &            & \multicolumn{1}{c}{$T_{\rm ex}$}
                      & \multicolumn{1}{c}{$\tau_{\nu_{\rm ul}}$}
                      & \multicolumn{1}{c|}{$N_{\rm tot}$}
                      &\multicolumn{1}{c}{$T_{\rm ex}$}
                      &\multicolumn{1}{c}{$\tau_{\nu_{\rm ul}}$}
                      &\multicolumn{1}{c}{$N_{\rm tot}$}  \\
         &            & \multicolumn{1}{c}{[K]}
                      &
                      & \multicolumn{1}{c|}{[cm$^{-2}$]}
                      &\multicolumn{1}{c}{[K]}
                      &                                                       
                      &\multicolumn{1}{c}{[cm$^{-2}$]} \\ \hline
                                                       
$^{34}$SO &$4_3 \rightarrow 3_2$& - & - &-
&$3.9\!\pm\!0.1$&$0.69\!\pm\!0.01$&$1.9\!\pm\!0.1\,\cdot 10^{13}$\\

SO&$4_3\rightarrow3_2$&$4.4\!\pm\!0.1$&$0.59\!\pm\!0.07$&
$2.0\!\pm\!0.2\,\cdot10^{13}$&$5.8\!\pm\!0.9$&$2.10\!\pm\!0.70$&$6.6\!\pm\!2.3\,\cdot
10^{13}$\\

SO&$3_3 \rightarrow 2_2$& - & - &-
&$5.8\!\pm\!0.9$&$0.37\!\pm\!0.05$&$7.2\!\pm\!2.2\,\cdot 10^{13}$\\

$^{34}$SO&$3_2 \rightarrow 2_1$& -& -
&-&$3.9\!\pm\!0.1$&$1.16\!\pm\!0.01$&$2.0\!\pm\!0.1\,\cdot 10^{13}$\\

SO&$3_2 \rightarrow
2_1$&$4.4\!\pm\!0.1$&$1.10\!\pm\!0.20$&$2.7\!\pm\!0.3\,\cdot
10^{13}$&$5.8\!\pm\!0.9$&$2.30\!\pm\!0.90$&$6.7\!\pm\!2.7\,\cdot
10^{13}$\\

C$^{34}$S&$3\rightarrow
2$&$3.3\!\pm\!0.1$&$0.49\!\pm\!0.04$&$5.7\!\pm\!0.5\,\cdot10^{12}$&
$3.4\!\pm\!0.1$&$0.38\!\pm\!0.04$&$3.9\!\pm\!0.5\,\cdot10^{12}$\\

C$^{34}$S&$2 \rightarrow 1$&$3.3\!\pm\!0.1$&$1.00\!\pm\!0.10$&
$5.7\!\pm\!0.6\,\cdot
10^{12}$&$3.4\!\pm\!0.1$&$0.83\!\pm\!0.12$&$5.0\!\pm\!0.7\,\cdot10^{12}$\\

CCS&$8_7\rightarrow7_6$&$4\!\pm\!1^{\rm a}$&-&$3.1\!\pm\!0.3\,\cdot
10^{13}$&$4\!\pm\!1^{\rm a}$&-&$4.0\!\pm\!2.0\,\cdot10^{12}$\\

o-c-C$_3$H$_2$&$3_{1,2}\rightarrow2_{2,1}$&$4.7\!\pm\!0.2$&
$0.83\!\pm\!0.14$&$4.7\!\pm\!0.8\,\cdot 10^{13}$&-&-&-\\

o-c-C$_3$H$_2$&$2_{1,2}\rightarrow1_{0,1}$&$4.7\!\pm\!0.2$&
$2.80\!\pm\!1.60$&$4.7\!\pm\!2.7\,\cdot 10^{13}$&$4\!\pm\!1^{\rm
a}$&-&$1.2\!\pm\!0.2\,\cdot 10^{12}$\\

CH$_3$OH&$3_{-1}\rightarrow2_{-1}\,E$&$4\!\pm\!1^{\rm a}$&-&
$6.8\!\pm\!1.3\,\cdot10^{13}$&$4\!\pm\!1^{\rm
a}$&-&$8.0\!\pm\!2.0\,\cdot10^{13}$\\

CH$_3$OH&$3_{0}\rightarrow2_{0}\,A^+$&$4\!\pm\!1^{\rm a}$&-&
$1.2\!\pm\!0.2\,\cdot10^{14}$&$4\!\pm\!1^{\rm a}$&-&$1.7\!\pm\!0.3\,\cdot10^{14}$\\

HN$^{13}$C&$1\rightarrow 0$&-&-&-&$4\!\pm\!1^{\rm
a}$&-&$3.2\!\pm\!0.7\,\cdot 10^{11}$ \\

HNC&$1\rightarrow 0$&-&-&-&$4\!\pm\!1^{\rm a}$&-&$2.0\!\pm\!0.1\,\cdot
10^{12}$\\

HC$_3$N&$11\rightarrow10$&$4.1\!\pm\!0.1$&$1.10\!\pm\!0.20$&$4.5\!\pm\!0.8\,\cdot
10^{14}$&-&-&-\\

HC$_3$N&$10\rightarrow9$&$4.1\!\pm\!0.1$&$3.00\!\pm\!2.00$&$4.5\!\pm\!2.5\,\cdot
10^{14}$&$6\!\pm\!1^{\rm a}$&-&$<5.7\,\cdot 10^{11\;{\rm b}}$\\

CH$_3$CCH&$8_0\rightarrow7_0$&$9.9\!\pm\!0.8$&$0.05\!\pm\!0.01$&$1.1\!\pm\!0.2\,\cdot
10^{14}$&-&-&-\\

CH$_3$CCH&$8_1\rightarrow7_1$&$8.5\!\pm\!0.8$&$0.05\!\pm\!0.01$&$1.0\!\pm\!0.2\,\cdot
10^{14}$&-&-&-\\

CH$_3$CCH&$5_0\rightarrow4_0$&$9.9\!\pm\!0.8$&$0.09\!\pm\!0.01$&$1.1\!\pm\!0.2\,\cdot
10^{14}$&-&-&-\\

CH$_3$CCH&$5_1\rightarrow4_1$&$8.5\!\pm\!0.8$&$0.10\!\pm\!0.01$&$1.0\!\pm\!0.2\,\cdot
10^{14}$&-&-&-\\

N$_2$H$^+$&$1\rightarrow 0$&$4\!\pm\!1^{\rm a}$& - &
$1.4\!\pm\!0.1\,\cdot 10^{13}$&$4\!\pm\!1^{\rm
a}$&-&$2.6\!\pm\!0.1\,\cdot 10^{12}$\\

\hline
\multicolumn{8}{l}{$^{\rm a}$ The excitation temperature has been assumed.} \\
\multicolumn{8}{l}{$^{\rm b}$ The rms is so large compared to the line
intensity  that only a upper limit for the columndensity is given here.}
\end{tabular}
\label{table:columndensities}
\end{table*}

\clearpage

\begin{table}
\caption[C18Oratios]{Column densities relative
to C$^{18}$O$^*$.}
\begin{tabular}{l|c|c} \hline
Molecules & \multicolumn{1}{|c|}{Cha-MMS1} &
\multicolumn{1}{|c}{CrA~C} \\ 
 &  \multicolumn{1}{c|}{$N_{\rm tot}/N_{\rm C^{18}O}\;(\times 10^{3})$}&\multicolumn{1}{c}{$N_{\rm tot}/N_{\rm C^{18}O}\;(\times 10^{3})$}  \\ \hline

$^{34}$SO      &-              &$6.3\pm0.5$ \\
SO             &$8.0\pm1.0$    &$22\pm8\;\,$ \\
C$^{34}$S      &$2.3\pm0.3$    &$1.3\pm0.2$ \\
CCS            &$12\pm2\;\;$   &$1.3\pm0.7$ \\
o-c-C$_3$H$_2$ &$19\pm4\;\;$   &$< 0.4$ \\
CH$_3$OH       &$27\pm6\;\;$   &$27\pm7\;\,$ \\
HN$^{13}$C     &-              &$0.11\pm0.02$ \\
HNC            &-              &$0.67\pm0.06$ \\
HC$_3$N        &$\!\!\!180\pm35$      &$<0.2$ \\
CH$_3$CCH      &$44\pm9\;\;$   &- \\
N$_2$H$^+$     &$5.6\pm0.6$    &$0.87\pm0.07$ \\
\hline
\end{tabular}
\label{table:fractions}

$^*$ The C$^{18}$O/H$_2$ column density ratio is normally assumed to
be about $2\cdot10^{-7}$. In quiescent regions like CrA C this ratio can be,
however, lower by a factor of two (\cite{harjunpaa}).    

\end{table}

\begin{table}
\caption[fractions]{Column density ratios.}
\begin{tabular}{l|c|c} \hline
Molecules & \multicolumn{1}{|c|}{Cha-MMS1} &
\multicolumn{1}{|c}{CrA~C} \\ \hline
$\rm HC_3N\,/\,\rm CCS$      &$15\pm3\;\;$ &$< 0.14$ \\
$\rm HC_3N\,/\,\rm SO$       &$23\pm5\;\;$ &$< 0.01$ \\
$\rm HC_3N\,/\,\rm N_2H^+$   &$32\pm6\;\;$ &$< 0.22$ \\
$\rm HC_3N\,/\,\rm CH_3CCH$  &$\;\;4\pm1\;\;$ & - \\
$\rm HC_3N\,/\,\rm C_3H_2$   &$10\pm3\;\;$ &$< 0.48$ \\
$\rm C^{34}S\,/\,\rm ^{34}SO$&$6.4\pm0.9^{\rm a}$ &$0.21\pm0.03$ \\
$\rm N_2H^+\,/\,\rm SO$      &$0.7\pm0.1$ &$0.04\pm0.01$ \\
$\rm CCS\,/\,\rm N_2H^+$     &$2.2\pm0.3$ &$1.5\pm0.8$ \\
$\rm C^{34}S\,/\,\rm N_2H^+$ &$0.41\pm0.05$ &$1.5\pm0.2$\\
\hline
\multicolumn{3}{l}{$^{\rm a}$ The $^{32}$S/$^{34}$S isotope ratio is
assumed to be 22.5.}
\end{tabular}
\label{table:ratios}
\end{table}

\clearpage

\clearpage

\appendix

\section{Formulae used in the excitation temperature and column density
calculations}

The excitation temperatures ($T_{\rm ex}$) and molecular column
densities ($N_{\rm tot}$) were estimated using a local thermodynamic  equilibrium (LTE) approach in
the sense that we assume a uniformly excited and Boltzmann distributed
population of the energy levels for the molecule question.  However,
we refrain from making any assumptions concerning the (line-centre)
optical depths ($\tau_{\nu_{\rm ul}}$). Instead we derive an exact
equation (A5 below) for a two-level system, from which $T_{\rm ex}$
can be determined numerically (see \cite{nummelin} for an LTE
treatment of the excitation of molecules with numerous observed lines with no a priori
assumptions about the optical depths).  \\

\noindent By assuming that $\tau_{\nu_{\rm ul}}$ has a gaussian velocity 
distribution, the formal definition of the optical depth leads to the
following expression for its value at the line-centre:

\begin{eqnarray}
\label{tau1}
\tau_{\nu_{\rm ul}} &  = & 10^{-41}\,\sqrt{\frac{\ln 2}{\pi}}\,\frac{16 \pi^3}{3\,{\rm h}} 
\frac{\mu_{\rm el.}^2 S_{\rm ul} g(I,K) N_{\rm tot}}{Q(T_{\rm ex})\Delta v} \times
\nonumber \\
& & \times \exp(-E_{\rm u}/{\rm k}\,T_{\rm ex})\,\left(\exp({\rm h}\,\nu_{\rm ul}/{\rm k}\,T_{\rm ex})-1\right)
\,  .
\end{eqnarray}

\noindent
The factor $10^{-41}$ stems from using Debye and km\,s$^{-1}$ as the
units for the electric dipole moment ($\mu_{\rm el.}$) and the FWHM
line-width ($\Delta v$), respectively.  Quantities related to the
observed transition are: the rest-frequency $\nu_{\rm ul}$ [Hz], the
line-strength $S_{\rm ul}$, and the energy of the upper level
$E_{\rm u}$. Furthermore, $N_{\rm tot}$ is the total column density
[cm$^{-2}$], $Q(T_{\rm ex})$ is the partition function, $g(I,K)$ is
the statistical weight due to possible nuclear spin and
$K$-degeneracy, and ${\rm h}$ and ${\rm k}$ are the Planck and Boltzmann constants
(given in cgs-units), respectively. \\

\noindent A second expression for $\tau_{\nu_{\rm ul}}$ is obtained from the ``antenna
equation'':

\begin{equation}
T_{\rm R} = \frac{{\rm h}\,\nu_{\rm ul}}{{\rm k}}
\left[F(\nu_{\rm ul},T_{\rm ex})-F(\nu_{\rm ul},T_{\rm bg})\right]
\left[1-\exp(-\tau_{\nu_{\rm ul}})\right]
\end{equation}

\noindent where

\begin{equation}
F(\nu_{\rm ul},T)=\frac{1}{\exp\left({\rm h}\,\nu_{\rm ul}/{\rm k}\,T\right)-1} \, .
\end{equation}

\noindent 
$T_{\rm B}$ is the brightness temperature calculated from the observed
$T_{\rm A}^*$ by correcting for relevant beam-efficiencies and
beam-filling factors.  In the analysis of the observational data, we
used the main-beam brightness temperature ($T_{\rm mb}$=$T_{\rm
A}^*/\eta_{\rm mb}$), assumed the beam-filling to be one for all
lines, and did not introduce any corrections due to different
beam sizes..  $T_{\rm bg}$ is the temperature of the background
radiation (here taken as a black-body at a temperature of 2.728\,K).\\

\noindent Eqs.(A2 and A3)  yield:

\begin{equation}
\label{tau2}
\tau_{\nu_{\rm ul}} = \ln \left[ \frac{\left[F(\nu_{\rm ul},T_{\rm ex})-F(\nu_{\rm ul},T_{\rm
bg}) 
\right]} {\left[F(\nu_{\rm ul},T_{\rm ex})-F(\nu_{\rm ul},T_{\rm bg})
\right]-\frac{{\rm k}\,T_{\rm B}}{{\rm h}\,\nu_{\rm ul}}}\right] \, .
\end{equation}

Combining the Eqs. (A1 and A4) and re-arranging the terms leads to an
expression, which besides $T_{\rm ex}$ contains only known constants
or observed quantities.  By applying this to two spectral lines of the
same molecule, and taking the ratio of the resulting expressions, the
final equation for $T_{\rm ex}$ reads:

\begin{eqnarray}
\label{artoTex2}
& \frac{  \ln \left[ \frac{\left[F(\nu_2,T_{\rm ex})-F(\nu_2,T_{\rm bg}) \right]}
{\left[F(\nu_2,T_{\rm ex})-F(\nu_2,T_{\rm bg}) \right]
-\frac{{\rm k}\,T_{\rm B,2}}{{\rm h}\,\nu_2}} \right]}
{  \ln \left[ \frac{\left[F(\nu_1,T_{\rm ex})-F(\nu_1,T_{\rm bg}) \right]}
{\left[F(\nu_1,T_{\rm ex})-F(\nu_1,T_{\rm bg}) \right]
-\frac{{\rm k}\,T_{\rm B,1}}{{\rm h}\,\nu_1}} \right]} 
\frac{F(\nu_2,T_{\rm ex})}{F(\nu_1,T_{\rm ex})}
\exp(\frac{E_{\rm u,2}-E_{\rm u,1}}{{\rm k}\,T_{\rm ex}}) & \nonumber \\
& = \frac{\Delta v_1 S_{\rm ul,2}}{\Delta v_2 S_{\rm ul,1}} \, . &
\end{eqnarray}

\noindent The subscripts ``1'' and ``2'' refer to ``line 1'' and
``line 2'', respectively. Terms involving $T_{\rm ex}$ have been 
collected on the left hand side of this expression.  We solved 
Eq.(A5) for given observed line parameters by searching through 
a sufficiently large $T_{\rm ex}$ interval. When applying this 
equation to the observed data, we found that in addition to the 
case with a single unique solution, there are
also cases where no solution exists, as well as cases where two
solutions are present.  The lack of solutions indicates that the
assumption of a uniform $T_{\rm ex}$ for both transitions has been
violated or that the beam-filling factors of the two transitions are
different; in this case we have chosen the point where we have the
least deviation from a solution as the value for $T_{\rm ex}$.  
The case with two solutions manifestates itself by giving one solution 
with a low $T_{\rm ex}$ and
high $\tau_{\nu_{\rm ul}}$, the second solution having a higher $T_{\rm
ex}$ and lower $\tau_{\nu_{\rm ul}}$. In the analysis we have chosen to 
use the low $T_{\rm ex}$ - high $\tau_{\nu_{\rm ul}}$ alternative.  \\

\noindent Once $T_{\rm ex}$ was estimated, $\tau_{\nu_{\rm ul}}$ was 
calculated from Eq.(A4) and subsequently $N_{\rm tot}$ was 
derived from Eq.(A1). \\

\noindent In cases where we have only one observed transition, 
we have assigned a $T_{\rm ex}$ close to that of a species with 
similar excitation requirements. $N_{\rm mol}$ is then obtained, 
assuming optically thin emission, from the formula:

\begin{equation}
N_{\rm tot}=\frac{1.67\cdot10^{23}\, 
\exp{\left(E_{\rm u}/{\rm k}\,T_{\mathrm{ex}}\right)}\, 
Q(T_{\mathrm{ex}})\, I_{\rm line}}{\,
\left(1-\frac{F(T_{\mathrm{bg}})}
{F(T_{\mathrm{ex}})}\right)\, \nu\, \mu_{\mathrm{el.}}^2\, 
S_{\rm ul}\, g(I,K)} \, .
\end{equation}

\noindent Here \mbox{$I_{\rm line}$=$\int\,T_{\rm B}\,{\rm d}v$} 
is the velocity-integrated line-intensity [K\,km\,s$^{-1}$]. \\

In the calculation of the column densities the partition functions
need to be evaluated.  For linear rotors ($^1\Sigma$) we used the
following formula (e.g. \cite{townes}, chapter 1):

\begin{equation}
Q\approx\frac{{\rm k}\,T_{\rm ex}}{{\rm h}\,B}+\frac{1}{3}
\end{equation}

\noindent where $B$ is the rotation constant. \\

\noindent For CCS and SO we made a direct 
summation over the energy levels with $E_{\rm i}/{\rm k}\, <575$\, K:

\begin{equation}
Q=\sum_i\, g_i\, \exp\left(-E_{\rm i}/{\rm k}\,T_{\rm ex}\right)
\end{equation}

\noindent where $g_i$ is the statistical weigth, $2J_i+1$, with the relevant $J_i$
used. \\

\noindent We would like to emphasize that the evaluation of the
partition function through direct summation is essential in the case
of SO and CCS. For example for SO, the ``usual''
integral approximation of the partition function
\mbox{($3\,\frac{{\rm k}\,T_{\rm ex}}{{\rm h}\,B}-1$)} overestimates $Q$ by a
factor of 1.1 to 1.85 in the temperature range 2--70\,K; the largest
discrepancy occurs at $T_{\rm ex}\approx5$\,K.  Thus for SO this
simple approximation can be used only for temperatures above 80\,K or
so.  A better analytical approximation for $Q$ at temperatures below
80\,K can be obtained by treating each of the three $N$ ladders as
separate rigid rotors and integrating over $J$. Note that only one of
the ladders ($J=N-1$) has $J$=0 as its lowest level; the two other
ladders begin with $J$=1.  The resulting partition functions are:

\begin{equation}
Q_{J=N-1, {\rm ladder}} \approx \frac{{\rm k}\,T_{\rm ex}}{{\rm h}\,B}+\frac{1}{3} \,
\end{equation}

\begin{equation}
Q_{J=N, {\rm ladder}} \approx \left(\frac{{\rm k}\,T_{\rm ex}}{{\rm h}\,B}-\frac{2}{3}\right)\, 
\exp(-\Delta E_1/kT_{\rm ex})\, 
\end{equation}

\begin{equation}
Q_{J=N+1, {\rm ladder}} \approx \left(\frac{{\rm k}\,T_{\rm
ex}}{{\rm h}\,B}-\frac{2}{3}\right)\, 
\exp(-\Delta E_2/kT_{\rm ex})\, 
\end{equation}

\noindent where $\Delta E_1$ is the energy difference 
between the lowest levels in the $J=N$ and $J=N-1$ ladders, and
$\Delta E_2$ the corresponding difference between the $J=N+1$ and
$J=N-1$ ladders. The lowest states in these ladders are $J_N$=$0_1$
(the ground-state), $1_1$, and $1_0$, respectively.  \\

\noindent The total partition function is then:

\begin{equation}
Q = Q_{J=N-1, {\rm ladder}} + Q_{J=N, {\rm ladder}} + Q_{J=N+1, {\rm ladder}} \, .
\end{equation}

\noindent For SO, this approximation overestimates the true partition 
function by less than 15\% for temperatures between 3 and 80\,K, the
error being less than 5\% for $T_{\rm ex}>25$\,K.  The values for 
$\Delta E_1$ and $\Delta E_2$ are 15.18\,K and 1.44\,K (SO); 
9.33\,K and 0.53\,K (CCS),
respectively. \\

\noindent For c-C$_3$H$_2$ we used:
\begin{equation}
Q\approx2\, \sqrt[]{\frac{\pi\,{\rm k}^{3}T_{\rm ex}^{3}}{{\rm h}^3\,ABC}}
\approx 2.450\,T_{\rm ex}^{3/2} \, .
\end{equation}

\noindent The contributions from both ortho- and para-states  
have been accounted for by taking into account their different 
statistical weigths: 3 for ortho and 1 for para.   \\

\noindent For CH$_3$OH we used:
\begin{equation}
Q\approx2\, \sqrt[]{\frac{\pi\,{\rm k}^{3}T_{\mathrm{ex}}^{3}}{{\rm h}^3\,ABC}}
\approx 1.232\, T_{\mathrm{ex}}^{3/2} \, .
\end{equation}

\noindent 
The (equal) contributions from both the $A$- and the
$E$-symmetry species have been accounted for. \\

\noindent For CH$_3$CCH it is common to use the expression: \\
\begin{equation}
Q\approx\frac{8}{3}\, \sqrt[]{\frac{\pi\,{\rm k}^3 T_{\rm ex}^3}{{\rm h}^3\,ABC}}
\approx 4.177\,T_{\rm ex}^{3/2} \, .
\end{equation}

where the contributions from both the $A$- and the $E$-symmetry
species have been accounted for by taking into consideration the spin
statistics of the various levels (see e.g. \cite{townes}, chapter
3). \\

However, Eq.(A15) is valid only at high enough temperatures.  Since
the lines observed by have low excitation temperatures, we derived a
more accurate approximation for temperatures below 30\,K.  Each
$K$-ladder was treated as a linear rotor and the integration was
performed over $J$.  Note that $J$=$K$ is the lowest level of a
specific $K$-ladder.  By taking into account only the three first
ladders ($K$=0, 1, 2) we have:

\begin{equation}
Q_{K=0, {\rm ladder}} \approx \frac{{\rm k}\,T_{{\rm ex},J}}{{\rm h}\,B}+\frac{4}{3} \,
\end{equation}

\begin{equation}
Q_{K=1, {\rm ladder}} \approx \left(\frac{{\rm k}\,T_{{\rm
ex},J}}{{\rm h}\,B}-\frac{8}{3}\right)\, 
\exp(-\Delta E_3/kT_{{\rm ex},K})\,
\end{equation}

\begin{eqnarray}
Q_{K=2, {\rm ladder}} & \approx & 
\left(\frac{{\rm k}\,T_{{\rm ex},J}}{{\rm h}\,B}-\frac{8}{3}-12\,
\exp(-\Delta E_3/kT_{{\rm ex},J})\right)\times \nonumber \\
& & \times\exp(-\Delta E_4/kT_{{\rm ex},K})\,
\end{eqnarray}

\noindent where $\Delta E_3/{\rm k}\approx7.20$\,K and $\Delta E_4/{\rm k}\approx28.80$\,K are
the 
energy differences between the lowest levels of the $K$=1 and $K$=2
ladders, respectively, relative to the $K$=0 ladder.  $T_{{\rm ex},J}$
is the excitation temperature along a specific $K$-ladder (i.e. $K$
fix, $J$ varies), and $T_{{\rm ex},K}$ is the excitation temperature
between different $K$-ladders (i.e. $K$ varies, $J$ fix). \\

The resulting expression for the total partition function for
CH$_3$CCH, which in the indicated temperature range underestimates the
result from a direct summation by less than 10\%, is given by:

\begin{equation}
Q \approx Q_{K=0, {\rm ladder}} + Q_{K=1, {\rm ladder}} + Q_{K=2, {\rm ladder}}
\end{equation}


\clearpage

\begin{thebibliography}{}

\bibitem[Anderson et al. 1990]{anderson90} Anderson T., De Lucia F.C., 
         Herbst E. 1990, ApJS 72, 797

\bibitem[Benson et al. 1998]{benson98} Benson P.J., Caselli P., 
         Myers P.C. 1998, ApJ 506, 743

\bibitem[Bergin \& Langer 1997]{bergin97} Bergin E.A., Langer W.D. 
         1997, ApJ 486, 316

\bibitem[Bergin et al. 1994]{bergin94} Bergin E.A., Goldsmith P.F., 
         Snell R.L., Ungerechts H. 1994, ApJ 431, 674

\bibitem[Caselli et al. 1995]{caselli95}Caselli P., Myers P.C., 
         Thaddeus P. 1995, ApJ 455, L77

\bibitem[Ciolek \& Mouschovias 1993]{ciolek93} Ciolek G.E., Mouschovias T.C. 1993, 
        ApJ 418, 774

\bibitem[Ciolek \& Mouschovias 1994]{ciolek94} Ciolek G.E., Mouschovias T.C. 1994, 
        ApJ 425, 142

\bibitem[Ciolek \& Mouschovias 1995]{ciolek95} Ciolek G.E., Mouschovias T.C. 1995, 
        ApJ 454, 194
  
\bibitem[Cox et al. 1989]{cox} Cox P., Walmsley C.M., G\"usten R. 1989, 
         A\&A 209, 382

\bibitem[Gordy \& Cook 1970]{gordy} Gordy W., Cook R.L. 1970, {\sl Microwave 
         Molecular Spectra}, John Wiley \& Sons, Inc., New York 

\bibitem[Federman et al. 1990]{federman} Federman S.R., Huntress W.T., Jr., 
         Prasad S.S. 1990, ApJ 354, 504 

\bibitem[Frerking et al. 1979]{frerking} Frerking M.A., Langer W.D., 
         Wilson R.W. 1979, ApJ 232, L65

\bibitem[Harju et al. 1993]{harju93} Harju J., Haikala L.K., Mattila K., 
         et al. 1993, A\&A 278, 569

\bibitem[Harjunp\"a\"a \& Mattila 1996]{harjunpaa} Harjunp\"a\"a P., Mattila
         K. 1996, A\&A 305, 920

\bibitem[Hartquist \& Williams 1989]{hartquist89} Hartquist T.W.,
Williams D.A. 1989, MNRAS 241, 417
 
\bibitem[Hartquist et al. 1996]{hartquist96} Hartquist T.W., Williams D.A., 
Caselli P. 1996, Ap\&SS 238, 303

\bibitem[Herbst 1978]{herbst78} Herbst E. 1978, ApJ 222, 508

\bibitem[Herbst \& Leung 1989]{herbst89} Herbst E., Leung C.M. 1989, ApJS 69, 271

\bibitem[Hirahara et al. 1992]{hirahara92} Hirahara Y., Suzuki H., 
         Yamamoto S., et al. 1992, ApJ 394, 539

\bibitem[Hirahara et al. 1995]{hirahara95} Hirahara Y., Masuda A., 
         Kawaguchi K., et al. 1995, PASJ 47, 1

\bibitem[Huntress \& Mitchell 1979]{huntress} Huntress W.T. Jr.,  Mitchell
         G.F. 1979, ApJ 231, 456

\bibitem[Knude \& H{\o}g 1998]{knude}Knude J., H{\o}g E. 1998, A\&A 338, 897

\bibitem[Lehtinen et al. 2000]{lehtinen} Lehtinen K., Haikala L.K.,
         Mattila K., Lemke D. 2000, {\sl submitted to} A\&A

\bibitem[Little et al. 1979]{little} Little L.T., MacDonald G.H., Riley
         P.W., Matheson D.N. 1979, MNRAS 189, 539

\bibitem[Lizano \& Shu 1987]{lizano} Lizano S., Shu F.H. 1987, in {\sl
        Physical Processes in Interstellar Clouds}, Eds. Morfill G.E., Scholer
        M., D.Reidel, Dordrecht, Holland, p. 173

\bibitem[Llewellyn et al. 1981]{llewellyn} Llewellyn R., Payne P., Sakellis S., 
         Taylor K.N.R., 1981, MNRAS 196, 29P

\bibitem[Mattila et al. 1989]{mattila} Mattila K., Liljestr\"om T., Toriseva M. 1989, 
         in {\sl Low Mass Star Formation and Pre-Main Sequence Objects}, Reipurth B. (ed.),
         ESO Conference and Workshop Proceedings 33, p. 153 

\bibitem[Menten et al. 1988]{menten} Menten K.M., Walmsley C.M., Henkel C.,
         Wilson T.L. 1988, A\&A 198, 253 

\bibitem[Mouschovias 1979]{mouschovias79} Mouschovias T.C. 1979, ApJ 228, 475

\bibitem[Murukami 1990]{murukami} Murukami A. 1990, ApJ 357, 288

\bibitem[Myers \& Benson 1983]{myers83} Myers P.C., Benson P.J. 1983, 
         ApJ 266, 309

\bibitem[Nejad \& Wagenblast 1999]{nejad} Nejad L.A.M., Wagenblast
         R. 1999, A\&A 350, 204

\bibitem[Nilsson et al. 2000]{nilsson} Nilsson A., Hjalmarson {\AA}., 
         Bergman P., Millar T.J. 2000, A\&A {\it in press}  

\bibitem[Nummelin et al. 2000]{nummelin} Nummelin A., Bergman P., Hjalmarson {\AA}., 
         Friberg P., Irvine W.M. 2000, ApJ {\it in press}   

\bibitem[Poynter \& Pickett 1985]{poynter} Poynter R.L., Pickett
         H.M. 1985, Appl. Opt. 24, 2335

\bibitem[Prasad \& Huntress 1982]{prasad} Prasad S.S., Huntress W.T., Jr. 1982, ApJ 260, 590

\bibitem[Pratap et al. 1997]{pratap} Pratap P., Dickens J. E., Snell
         R. L., et al. 1997, ApJ 486, 862

\bibitem[Prusti et al. 1991]{prusti} Prusti T., Clark F.O., Whittet D.C.B., 
         Laurejs R.J., Zhang C.Y. 1991, MNRAS 251, 303

\bibitem[Rawlings et al. 1992]{rawlings} Rawlings J.M.C., Hartquist T.W.,
         Menten K.M., Williams D.A. 1992, MNRAS 255, 471

\bibitem[Reipurth et al. 1996]{reipurth} Reipurth B., Nyman L.-\AA., Chini R. 
         1996, A\&A 314, 258

\bibitem[Ruffle et al. 1997]{ruffle97} Ruffle D.P., Hartquist T.W., 
         Taylor S.D., Williams D.A. 1997, MNRAS 291, 235

\bibitem[Ruffle et al. 1999]{ruffle99} Ruffle D.P., Hartquist T.W., 
         Caselli P., Williams D.A. 1999, MNRAS 306, 691

\bibitem[Sastry et al. 1981]{sastry} Sastry K.V.N.L., Lees R.M., 
         Van der linde J. 1981, J. Mol. Spec 88, 228

\bibitem[Spitzer 1978]{spitzer} Spitzer, L. Jr. 1978, {\sl Physical
Processes in the Interstellar Medium}, John Wiley \& Sons, New York,
Ch. 13.3

\bibitem[Suzuki et al. 1992]{suzuki} Suzuki H., Yamamoto S., Ohishi M., 
         et al. 1992, ApJ 392, 551

\bibitem[Swade 1989]{swade} Swade D.A. 1989, ApJ 345, 828

\bibitem[Talbi \& Herbst 1998]{talbi98} Talbi D., Herbst E. 1998, A\&A
333, 1007

\bibitem[Townes \& Schawlow 1975]{townes} Townes C.H., Schawlow A.L. 1975,
         {\sl Microwave Spectroscopy}, Dover Publications Inc., New York

\bibitem[Turner et al. 1998]{turner98} Turner B.E., Lee H.-H., Herbst E.
         1998, ApJS 115, 91

\bibitem[Turner et al. 1999]{turner99} Turner B.E., Terzieva R., Herbst E.
         1999, ApJ 518, 699
         
\bibitem[van Dishoeck \& Blake 1998]{dishoeck} van Dishoeck E.F.,
         Blake G.A. 1998, ARA\&A 36, 317

\bibitem[Vanhala \& Cameron 1997]{vanhala} Vanhala H.A.T., Cameron
         A.G.W. 1997, ApJ 508, 291

\bibitem[Vrtilek et al. 1987]{vrtilek}Vrtilek J.M., Gottlieb C.A.,
         Thaddeus P. 1987, ApJ 314, 716

\bibitem[Yamamoto et al. 1990]{yamamoto}Yamamoto S., Saito S.,
         Kawaguchi K., et al. 1990, ApJ 361, 318 

\end{thebibliography}
\end{document}